\documentclass[12pt,twoside]{article}

\usepackage[
  left=3cm,right=2.5cm,top=3.1cm,bottom=2cm,headheight=2cm,headsep=1.2cm,footskip=1.4cm,twoside
]{geometry}
\usepackage[margin=1cm,font=small]{caption}
\usepackage{amsmath,amssymb}
\usepackage{lineno}
\usepackage{url}
\usepackage{subfigure}
\usepackage{pstricks}
\newcmykcolor{darkgreen}{1 0 0.6 0.5}  

\usepackage{epsfig}
\usepackage{epsf}
\usepackage{cite}   

\newlength{\nseparation}
\input{babarsymMerge-chka}

\usepackage{cite}   

\newcommand\Amptpbar{\kern 0.18em\overline{\kern -0.18em {\cal A}}_{3\pi}}

\newcommand\Amptpbarkappa{\kern 0.18em\overline{\kern -0.18em A}^{\kappa}{}}
\newcommand\Amptpbarsigma{\kern 0.18em\overline{\kern -0.18em A}^{\sigma}{}}

\newcommand\Nbpm{{\kern 0.18em\overline{\kern -0.18em N}}^{+-}}
\newcommand\Nbmp{{\kern 0.18em\overline{\kern -0.18em N}}^{-+}}

\newcommand\BRpmb{{\cal \kern 0.18em\overline{\kern -0.18em  B}}{}_{\rho\pi}^{+-}}
\newcommand\BRmpb{{\cal \kern 0.18em\overline{\kern -0.18em  B}}{}_{\rho\pi}^{-+}}

\newcommand\BRipmb{{\cal \kern 0.18em\overline{\kern -0.18em  B}}{}_{\rho^+\pi^-}}
\newcommand\BRimpb{{\cal \kern 0.18em\overline{\kern -0.18em  B}}{}_{\rho^-\pi^+}}

\newcommand\Abar{\kern 0.18em\overline{\kern -0.18em A}{}}

\def\journalL#1#2#3#4#5{\journal{#1 #2}{#3}{#4}{#5}}
\def\journal#1#2#3#4{#1~{\bf #2}, #3 (#4)}

\def\PRD#1#2#3{\journal{Phys.\ Rev. D}{#1}{#2}{#3}}
\def\PRL#1#2#3{\journal{Phys.\ Rev. Lett.}{#1}{#2}{#3}}

\newcommand{\etal}{\emph{et al.}\xspace}

\begin{document}
\begin{titlepage}
  {\raggedleft HU-EP-10/10 \\
  }
  \vskip 2em
  {\centering
    \large 
    H.~Lacker and A. Menzel [CKMfitter Group]\\
  }
  \vskip 2em
  {\centering {\Huge Simultaneous Extraction\\ 
                     of the Fermi constant\\ 
                     and PMNS matrix elements\\
                     in the presence \\
                     of a fourth generation\\}}
  \vskip 2.5em
  {
    \noindent
    Several recent studies performed on constraints of a fourth 
    generation of quarks and leptons suffer from the ad-hoc assumption 
    that $3 \times 3$ unitarity holds for the first three generations in 
    the neutrino sector. Only under this assumption one is able to determine 
    the Fermi constant $G_F$ from the muon lifetime measurement with the 
    claimed precision of $G_F = 1.16637 (1) \times 10^{-5}~{\rm GeV}^{-2}$. 
    We study how well $G_F$ can be extracted within the framework of four 
    generations from leptonic and radiative $\mu$ and $\tau$ decays, as well 
    as from $K_{\ell 3}$ decays and leptonic decays of charged pions, and we 
    discuss the role of lepton universality tests in this context.
    We emphasize that constraints on a fourth generation from quark and 
    lepton flavour observables and from electroweak precision observables 
    can only be obtained in a consistent way if these three sectors are 
    considered simultaneously. In the combined fit to leptonic and radiative 
    $\mu$ and $\tau$ decays, $K_{\ell 3}$ decays and leptonic decays of charged 
    pions we find a p-value of $2.6~\%$ for the fourth generation
    matrix element $|U_{e 4}|=0$ of the neutrino mixing matrix. 
    }
    \vskip 2em \noindent
{\small \em Humboldt-Universit\"at zu Berlin,
                   Institut f\"ur Physik,
                   Newtonstr. 15,
                   D-12489 Berlin, Germany,
                {e-mail: lacker@physik.hu-berlin.de}}\\[0.2cm]
\end{titlepage}


\vspace*{1cm}

\section{Introduction}
After a rather long period of low activity concerning studies of an extension 
of the three generation Standard Model (3SM) to a fourth generation Standard 
Model (4SM) a revival of the topic is observed in recent years in the literature. 
The possible role of a fourth generation in explaining some interesting hints 
for deviations from the Standard Model (SM) in flavour physics has been 
discussed in several 
papers~\cite{fourthfamilyCPphaseBtoSArhrhribHou,fourthfamilyCPphaseBtoSandDmixingHouNagashimaSodu,fourthfamilyCPphaseKpinunubarKPiHouNagashimaSodu,fourthfamilyCPphaseKPiNLOPQCDHouLiMishimaNagashima,fourthfamilyBCPAsymmetriesSoniAlokGiriMohantaNandi,fourthfamilyBotellaBrancoNebot,Bobrowski,Soni:2010xh,Buras:2010pi}.
The electroweak precision fit with respect to a fourth generation has been 
reconsidered as well~\cite{HePolonskySu,Plehn,Vysotsky}. This triggered more 
intensive work on the fourth generation, e.g. on constraints from the electroweak 
precision fit~\cite{Chanowitz} taking into account also the dependence on CKM~\cite{ckm} 
matrix elements in the T-parameter, or on the role in the anomalous magnetic 
moment of the muon~\cite{HouMa}.
 
The above mentionned studies rely on the implicit, but often not recognized, 
assumption that the extraction of a central parameter in the SM is unaffected 
by the presence of a fourth generation: the value of the Fermi constant $G_F$ 
extracted from the lifetime measurement of muons. The assumption made is that 
the 3SM neutrino mixing (PMNS~\cite{PMNS}) matrix fulfills exact $3 \times 3$ 
unitarity already in the first three generations. Under this assumption the 
PMNS matrix elements $U_{e 4}$, $U_{\mu 4}$ and $U_{\tau 4}$ are all zero and 
the relative uncertainty of the extracted $G_F$ value ($10^{-5}$) is so small 
that all these studies simply consider $G_F$ as a parameter without uncertainties. 
One might naively expect small or even tiny PMNS elements for the mixing between 
a fourth generation neutrino, which must be heavy to escape the bound from the 
$Z$-resonance width~\cite{LEPZwidth}, and the light neutrinos of the first three 
generations. Nevertheless, the assumption of zero mixing elements is certainly 
not justified, in particular, since we neither understand how family replication 
works, nor how the values of the mixing matrix elements are fixed by Nature.

In this paper, we emphasize that taking $G_{F}$ as a constant parameter is not 
{\it a-priori} justified and that the precision on $G_F$ extracted from leptonic 
$\mu$ and $\tau$ decays and from $K_{\ell3}$ decays and leptonic decays of 
charged pions, once this assumption is abandoned, is only of order $50~\%$ 
despite the fact that lepton universality is tested to a very good precision 
in these decays. 
When adding the information from the search limits for the radiative decays 
$\mu \rightarrow e \gamma$, $\tau \rightarrow e \gamma$ and 
$\tau \rightarrow \mu \gamma$ we show that the precision on $G_F$ is 
significantly improved. Nevertheless, the precision of the extracted $G_F$ 
value is still about two orders of magnitude worse than in the 3SM case. 
As a consequence, statements made about a fourth generation scenario with 
respect to PMNS and/or CKM quark mixing matrix elements as well as concerning 
fourth generation quark and lepton mass differences and Higgs mass bounds from 
an electroweak precision fit like in~\cite{HePolonskySu,Plehn,Vysotsky} possibly
suffer from this source of uncertainty that has not been considered in these kind
of studies.\\
Studies of a non-unitary PMNS and/or CKM matrix are not completely new. Lepton 
or quark universality violations and possible effects on the extraction of $G_{F}$ 
has been already considered in several 
papers~\cite{LeeShrock,ShrockWang,ShrockTreimanWang,ShrockI,ShrockII,AppelquistPiaiShrock}. 
Mixing between ordinary and ``exotic'' fermions as an extension of the 3SM has 
been studied in quite some quantitative detail in Ref.~\cite{LangackerLondon}. 
Ref.~\cite{TammasiniBarenboimBernabeuJarlskog} focused on the aspects of the 
non-decoupling quadratic dependence on the heavy neutrino mass in lepton 
flavour violating processes like $\mu \rightarrow e \gamma$ and 
$\mu \rightarrow e e^{+} e^{-}$. In Ref.~\cite{AntuschBiggioFernandez-MartinezGavelaLopez-Pavon} 
a detailed analysis of possible violations of unitarity of the $3 \times 3$ PMNS 
matrix has been presented under the assumption of ``Minimal Unitarity 
Violation'' (MVU): only three light neutrinos are considered and New Physics is 
introduced in the neutrino sector only. The authors of 
Ref.~\cite{AntuschBiggioFernandez-MartinezGavelaLopez-Pavon} considered $W$ decays, 
invisible $Z$ decays, leptonic $\mu$ and $\tau$ decays as well as leptonic $\pi$ 
decays together with radiative $\mu$ and $\tau$ decays and combined this information 
for the first time with neutrino oscillation data. Refinements of this analysis 
can be found in Ref.~\cite{AntuschBaumannFernandez-Martinez} where also $G_F$ has 
been constrained using the unitarity constraint in the first row of the CKM matrix
while replacing the constraint from invisible $Z$ decays. 
Effects of a non-unitary PMNS matrix on CP violation in neutrino oscillation 
observables within the MVU framework have been studied in
Refs.~\cite{Fernandez-MartinezGavelaJLopez-PavonYasuda} 
and~\cite{AntuschBlennowFernandez-MartinezLopez-Pavon}.
\\ 
In our paper we explicitely study constraints when allowing for an additional 
fourth lepton and quark generation. Our analysis is contained in the more general 
analysis of Refs.~\cite{AntuschBiggioFernandez-MartinezGavelaLopez-Pavon} and 
\cite{AntuschBaumannFernandez-Martinez} but more specific since a smaller number 
of new parameters is needed to account for unitarity violations. The only case 
where the analysis of Ref.~\cite{AntuschBiggioFernandez-MartinezGavelaLopez-Pavon} 
is more specific than ours is the usage of invisible $Z$ decays where the authors 
of Ref.~\cite{AntuschBiggioFernandez-MartinezGavelaLopez-Pavon} only take into 
account those radiative corrections which originate from the 3SM.\\
The numerical studies presented in our paper have been performed using the 
software package CKMfitter~\cite{CKMfitter} where the four generation PMNS
and CKM matrix have been implemented using the Botella-Chau 
parametrisation~\cite{BotellaChau}.

\section{Extraction of $G_F$ from leptonic $\mu$ and $\tau$ decays}
Let us first consider the $\mu$ decay which is dominated by the final state
$\mu^{-} \rightarrow e^{-} {\bar{\nu}}_{e} \nu_{\mu}$. 
In the $\mu$-lifetime measurement neither the neutrinos in the final state 
are detected nor is it possible to decide for a given event which neutrino 
mass eigenstate has been produced. The predicted muon decay-rate 
$\Gamma(\mu \rightarrow all) = \Gamma(\mu^{-} \rightarrow e^{-} {\bar{\nu}}_{e} \nu_{\mu}) + \Gamma(\mu^{-} \rightarrow e^{-} {\bar{\nu}}_{e} \nu_{\mu} + \gamma) + ...$, which takes into account electroweak 
corrections~\cite{KinoshitaSirlin,Sirlin1,Marciano,RitbergenStuart}, 
is given by 
\begin{eqnarray}\label{Eq:MuDecayRate}
\Gamma(\mu^{-} \rightarrow all) &=&
\frac{G_{F}^{2} m_{\mu}^{5}}{192 \pi^{3}} \cdot PS(m_{e},m_{\mu}) \cdot \left[ (1-\frac{\alpha(m_{\mu}^{2})}{2\pi} (\pi^{2}-\frac{25}{4})+C_{2} \frac{\alpha^{2}(m_{\mu}^{2})}{\pi^2}) (1+\frac{3 m_{\mu}^{2}}{5 m_{W}^{2}})\right]\nonumber \\
&\cdot& \sum_{i=1,2,3} |U_{e i}|^{2} \sum_{j=1,2,3} |U_{\mu j}|^{2},
\end{eqnarray}
where $G_{F}$ is the Fermi constant to be determined, $m_{\mu}=(105.658367 \pm 0.000004)~{\rm MeV}$~\cite{PDG2008} 
and $m_{e}=(0.510998910 \pm 0.000000013)~{\rm MeV}$~\cite{PDG2008} is the muon, 
respectively, the electron mass, $m_{W}=(80.398 \pm 0.025)~{\rm MeV}$~\cite{PDG2008} 
is the $W$-boson mass, $\alpha(m_{\mu}^{2})$ is the fine structure constant 
at the scale $m_{\mu}^{2}$ calculated using Ref.~\cite{alphaqedjetset},
$C_{2}=(6.700 \pm 0.002)$ is the coefficient of the $O(\alpha^{2})$ radiative
corrections~\cite{RitbergenStuart}, and 
\begin{equation}\label{Eq:PhaseSpace}
PS(m_{l},m_{i}) = 
1 - 8 \left(\frac{m_{l}}{m_{i}}\right)^{2} 
  - 12 \left(\frac{m_{l}}{m_{i}}\right)^{4} \log\left(\frac{m_{l}}{m_{i}}\right)^{2} 
  +  8 \left(\frac{m_{l}}{m_{i}}\right)^{6} - \left(\frac{m_{l}}{m_{i}}\right)^{8}
\end{equation}
is a factor that accounts for the correction of the phase space due to finite masses 
in the final state. Here, it is assumed that only the electron has a significant mass 
while the neutrino masses are neglected in the phase space correction factor. 
The world average of the muon life $\tau_{\mu}=\Gamma^{-1}(\mu^{-} \rightarrow all)$
reads $\tau_{\mu}=(2.197034 \pm 0.000021) \cdot 10^{-6}~{\rm s}$~\cite{PDG2008}
where the uncertainty has been halved recently by measurements from the MuLan~\cite{MuLan} 
and the FAST~\cite{FAST} collaborations.

The $U_{e i}$ and $U_{\mu j}$ are the PMNS matrix elements for the three 
light neutrino mass states of the first three generations. 
In the 3SM the sums $\sum_{i=1,2,3} |U_{e i}|^{2}$ and $\sum_{j=1,2,3} |U_{\mu j}|^{2}$ 
are exactly equal to 1 due to $3 \times 3$ unitarity.
In the 4SM the fourth generationneutrino must have a mass of at least about half 
of the Z-boson mass as the number of light neutrino flavours from measuring 
the partial widths of visible final states and the total width of the 
Z-resonance at LEP does exclude a light fourth neutrino flavour~\cite{LEPZwidth}. 
Hence, the additional fourth generation neutrino is not kinematically accessible 
in $\mu$-decays. Although the same formula for the 
$\mu^{-} \rightarrow e^{-} {\bar{\nu}}_{e} \nu_{\mu}$ applies, the sums 
$\sum_{i=1,2,3} |U_{e i}|^{2}$ and $\sum_{j=1,2,3} |U_{\mu j}|^{2}$ do not 
necessarily add up to 1 any more. In a fourth generation scenario one can write
$\sum_{i=1,2,3} |U_{e i}|^{2}=1-|U_{e 4}|^{2}$ and 
$\sum_{j=1,2,3} |U_{\mu j}|^{2}=1-|U_{\mu 4}|^{2}$
thanks to $4 \times 4$ unitarity of the fourth generation PMNS matrix.

Similar in line the partial rates for the leptonic $\tau$ decays including 
electroweak corrections at leading logarithm are given by~\cite{Marciano,SwainTaylor} 
\begin{eqnarray}\label{Eq:TauLeptDecayRateE}
\Gamma(\tau^{-} \rightarrow e^{-} {\bar{\nu}}_{e} \nu_{\tau} + (\gamma))&=& 
\frac{G_{F}^{2} m_{\tau}^{5}}{192 \pi^{3}} \cdot PS(m_{e},m_{\tau}) \cdot \left[ (1-\frac{\alpha(m_{\tau}^{2})}{2\pi} (\pi^{2}-\frac{25}{4})) (1+\frac{3 m_{\tau}^{2}}{5 m_{W}^{2}})\right]\nonumber \\
&\cdot& (1-|U_{e 4}|^{2})(1-|U_{\tau 4}|^{2}),
\end{eqnarray}
and 
\begin{eqnarray}\label{Eq:TauLeptDecayRateMu}
\Gamma(\tau^{-} \rightarrow \mu^{-} {\bar{\nu}}_{\mu} \nu_{\tau} + (\gamma))&=& 
\frac{G_{F}^{2} m_{\tau}^{5}}{192 \pi^{3}} \cdot PS(m_{\mu},m_{\tau}) \cdot \left[ (1-\frac{\alpha(m_{\tau}^{2})}{2\pi} (\pi^{2}-\frac{25}{4})) (1+\frac{3 m_{\tau}^{2}}{5 m_{W}^{2}})\right]\nonumber \\
&\cdot& (1-|U_{\mu 4}|^{2})(1-|U_{\tau 4}|^{2}).
\end{eqnarray}
These decay rates are experimentally determined from the measured $\tau$ 
lifetime $\tau_{\tau} = (290.6 \pm 1.0) \cdot 10^{-15}~{\rm s}$~\cite{PDG2008} 
and the corresponding measured branching fractions as
$\Gamma(\tau^{-} \rightarrow e^{-} {\bar{\nu}}_{e} \nu_{\tau} + (\gamma)) = BF(\tau^{-} \rightarrow e^{-} {\bar{\nu}}_{e} \nu_{\tau} + (\gamma))/ \tau_{\tau}$ and
$\Gamma(\tau^{-} \rightarrow \mu^{-} {\bar{\nu}}_{\mu} \nu_{\tau} + (\gamma)) = BF(\tau^{-} \rightarrow \mu^{-} {\bar{\nu}}_{\mu} \nu_{\tau} + (\gamma))/ \tau_{\tau}$.
The dominant uncertainties in the prediction of these branching fractions are 
induced by the $\tau$ lifetime and the $\tau$ mass 
$m_{\tau}=(1776.84 \pm 0.17)~{\rm MeV}$~\cite{PDG2008}.

One might be tempted to assume that the well-measured lepton universality, 
when comparing the values for the Fermi constant extracted from the 
experimentally determined values of 
$\Gamma(\mu^{-} \rightarrow e^{-} {\bar{\nu}}_{e} \nu_{\mu})$,
$\Gamma(\tau^{-} \rightarrow e^{-} {\bar{\nu}}_{e} \nu_{\tau})$, and
$\Gamma(\tau^{-} \rightarrow \mu^{-} {\bar{\nu}}_{\mu} \nu_{\tau})$,
should constrain the elements $U_{e4}$, $U_{\mu 4}$ and $U_{\tau 4}$ 
to be close to zero. However, this is not the case since these measurements 
can only constrain the products $G_F^2 (1-|U_{\mu 4}|^{2})(1-|U_{e 4}|^{2})$,
$G_F^2 (1-|U_{\tau 4}|^{2})(1-|U_{e 4}|^{2})$, and
$G_F^2 (1-|U_{\mu 4}|^{2})(1-|U_{\tau 4}|^{2})$.
As a consequence, the observed lepton universality constrains only the ratios 
$|U_{e4}|/|U_{\mu 4}|$, $|U_{\tau 4}|/|U_{e 4}|$, and $|U_{\tau 4}|/|U_{\mu 4}|$ 
to stay close to 1. E.g., for sufficiently large $|U_{e4}|$ and $|U_{\mu 4}|$ 
values one has to a very good approximation
$(1-|U_{e 4}|^2)/(1-|U_{\mu 4}|^2)=1$ from which follows $|U_{e 4}|=|U_{\mu 4}|$. 
As a result, lepton universality alone allows surprisingly large values for 
$U_{e4}$, $U_{\mu 4}$ and $U_{\tau 4}$ and in this case these elements show 
a very strong correlation. The maximally allowed values are 
easily determined without any sophisticated numerical analysis: unitarity 
requires $|U_{e4}|^{2} + |U_{\mu 4}|^2 + |U_{\tau 4}|^2 \le 1$ while 
perfect universality results in $|U_{e4}|=|U_{\mu 4}|=|U_{\tau 4}|$. 
As a consequence, lepton universality is perfectly compatible with
$|U_{e4}|, |U_{\mu 4}|, |U_{\tau 4}| \le 1/\sqrt{3} \approx 0.577$. 
In this worst case limit, $G_F$ could have a value that is larger by a 
factor of $1.5$ with respect to the value extracted from the 3SM scenario.\\
In what follows we denote the value of $G_{F}$ extracted from the muon
lifetime measurement when assuming $3 \times 3$ unitarity of the PMNS 
matrix for the first three neutrino generations as 
$G_{F}^{3SM} = 1.16637 (1) \times 10^{-5}~{\rm GeV}^{-2}$ by setting 
$|U_{e4}|=|U_{\mu 4}|=|U_{\tau 4}|=0$ while in the general four 
generation case it is denoted as $G_{F}^{4SM}$.\\
\begin{figure}
     \centering
     \subfigure[]{
          \label{fig:a_Ue4Umu4Utau4}
          \includegraphics[width=.475\textwidth]{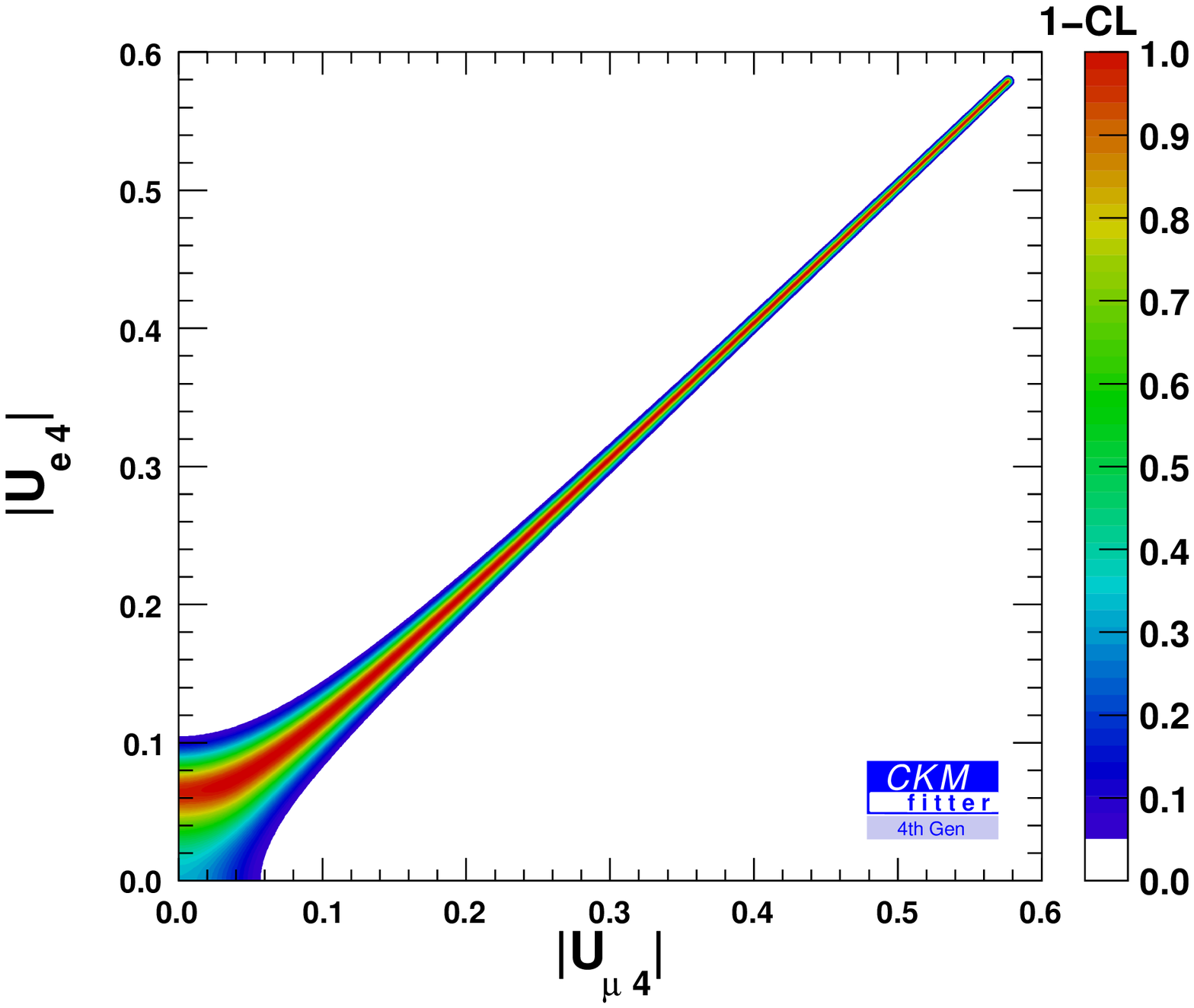}}
     \hspace{.01in}
     \subfigure[]{
          \label{fig:b_Ue4Umu4Utau4}
          \includegraphics[width=.475\textwidth]{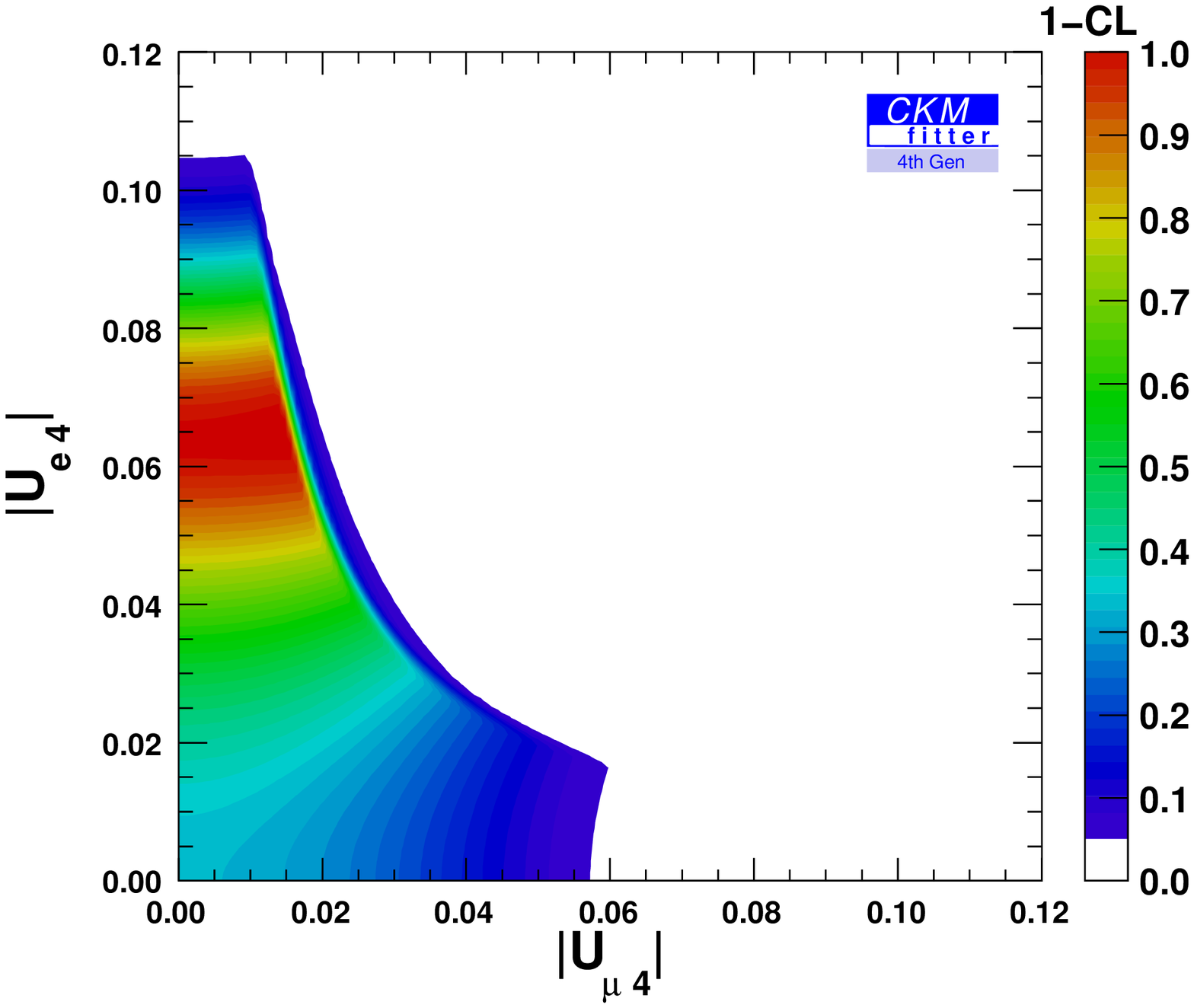}}\\
     \vspace{.01in}
     \subfigure[]{
          \label{fig:c_Ue4Umu4Utau4}
          \includegraphics[width=.475\textwidth]{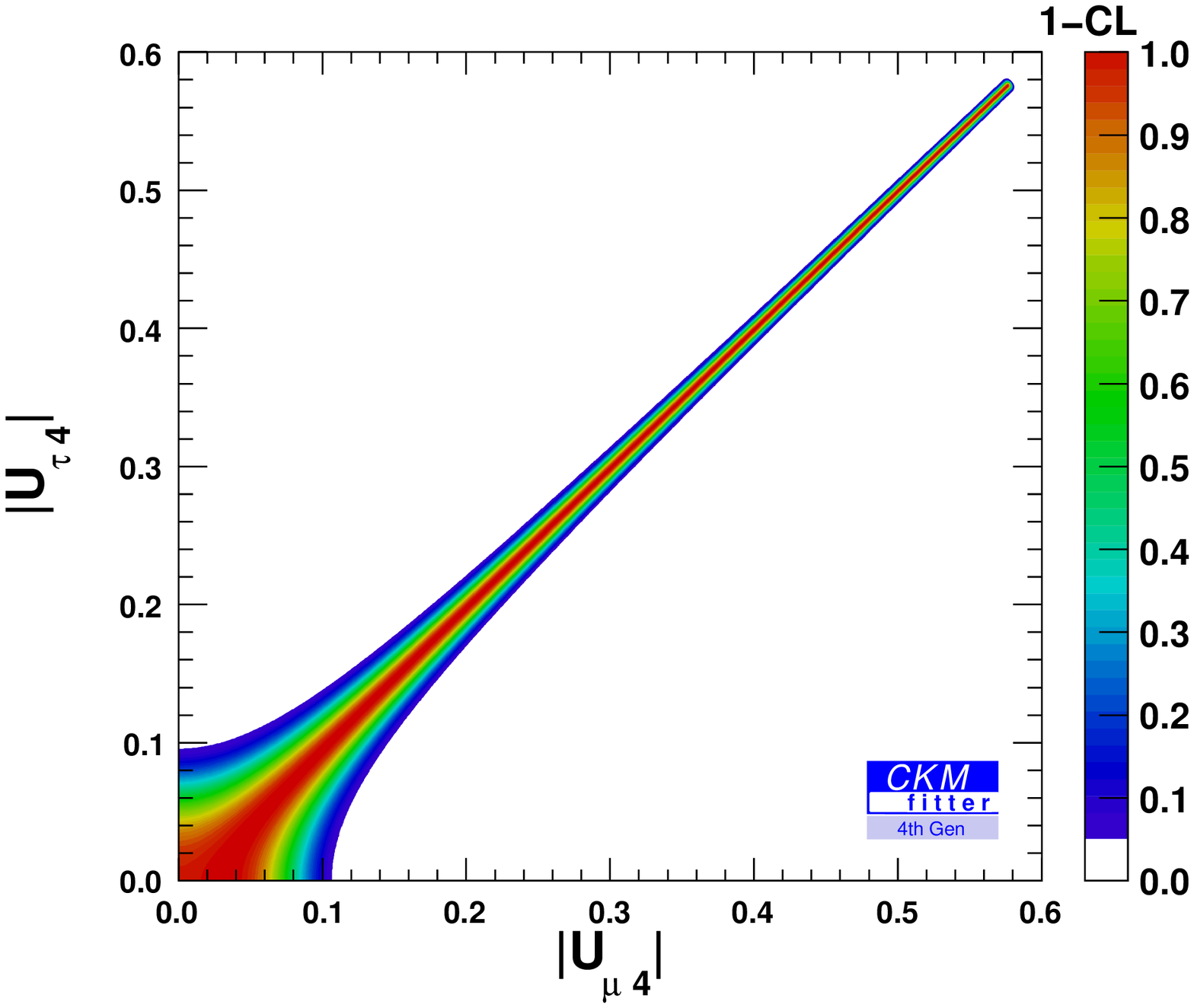}}
     \hspace{.01in}
     \subfigure[]{
          \label{fig:d_Ue4Umu4Utau4}
          \includegraphics[width=.475\textwidth]{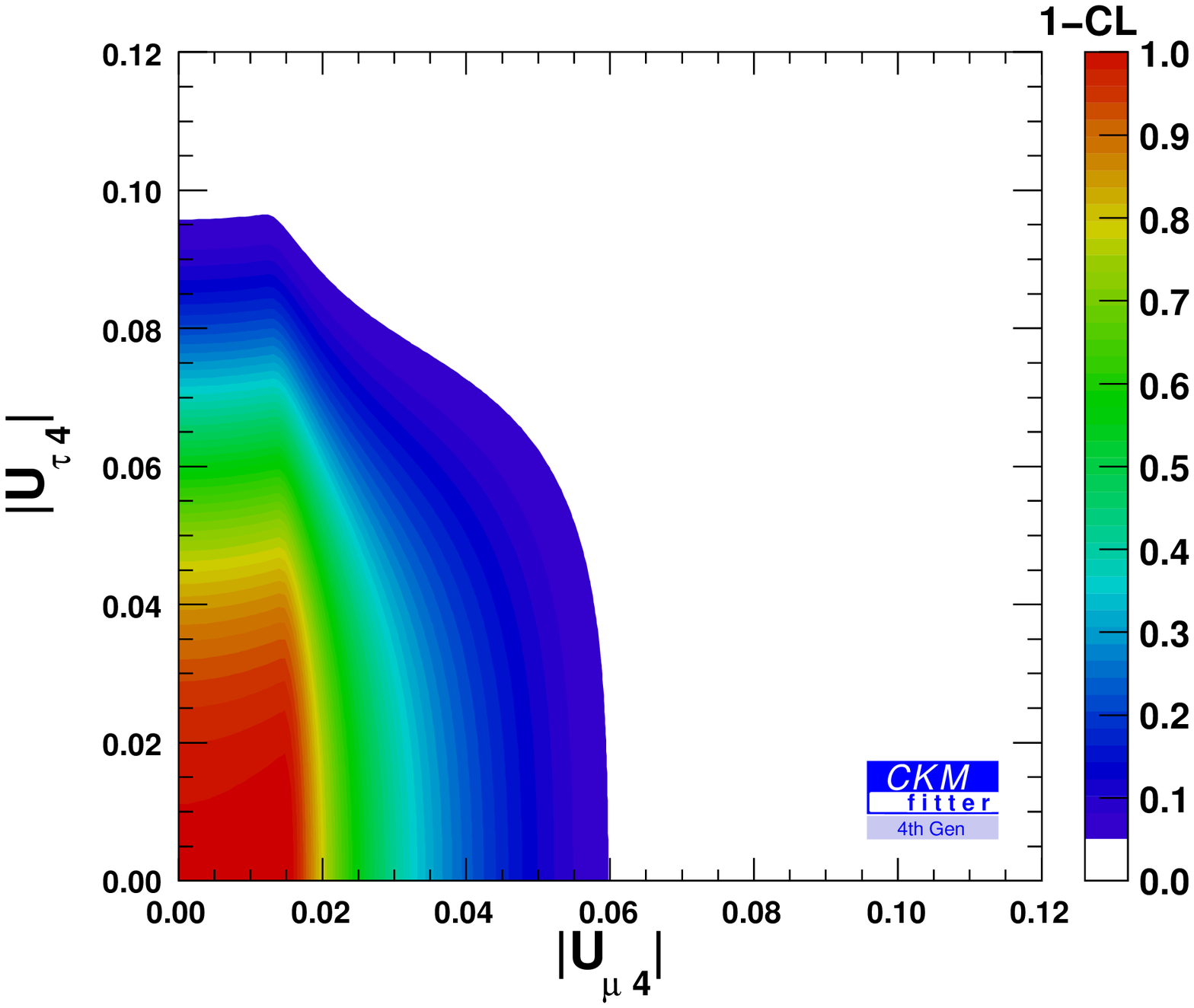}}\\
     \vspace{.01in}
     \subfigure[]{
          \label{fig:e_Ue4Umu4Utau4}
          \includegraphics[width=.475\textwidth]{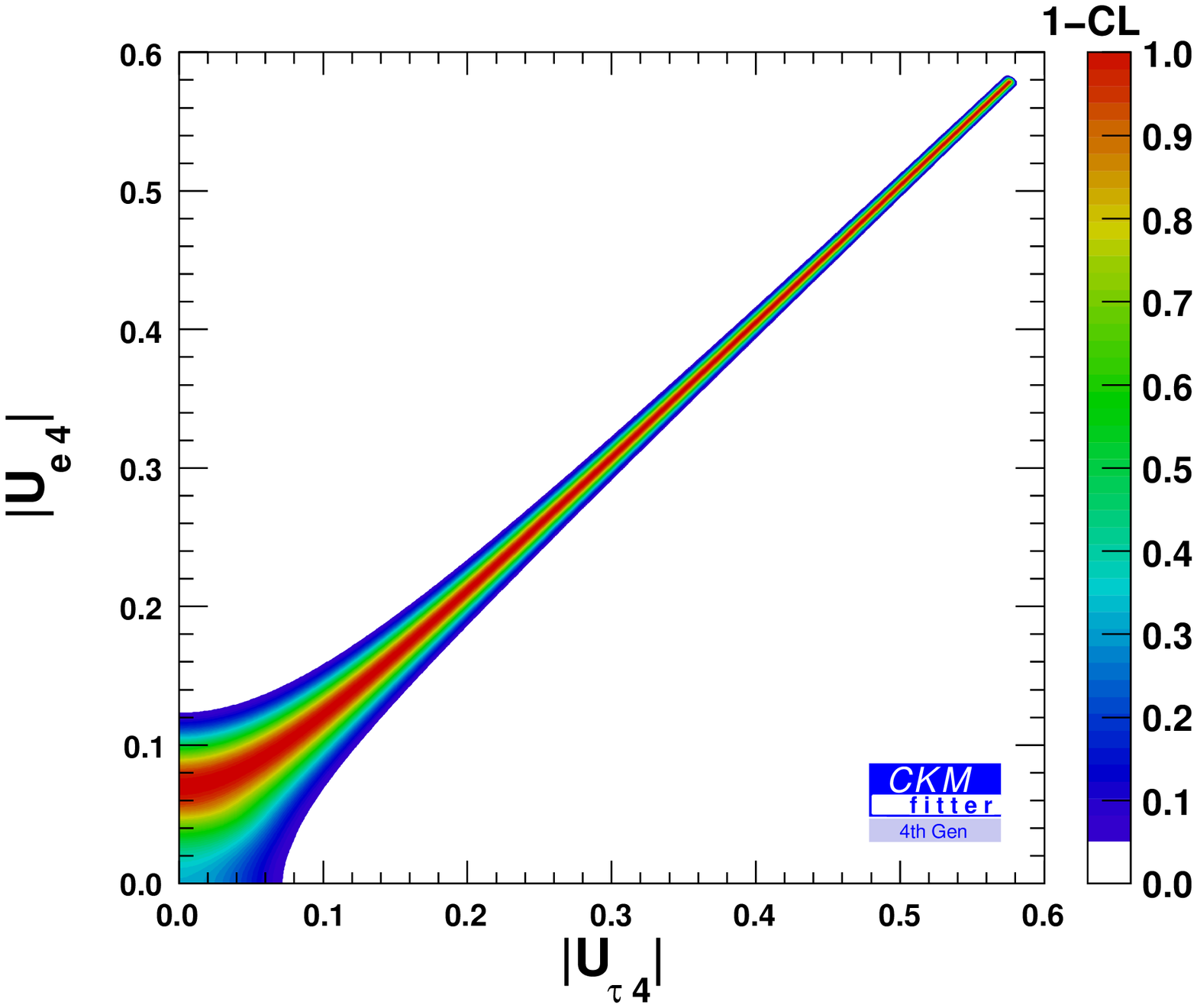}}
     \hspace{.01in}
     \subfigure[]{
          \label{fig:f_Ue4Umu4Utau4}
          \includegraphics[width=.475\textwidth]{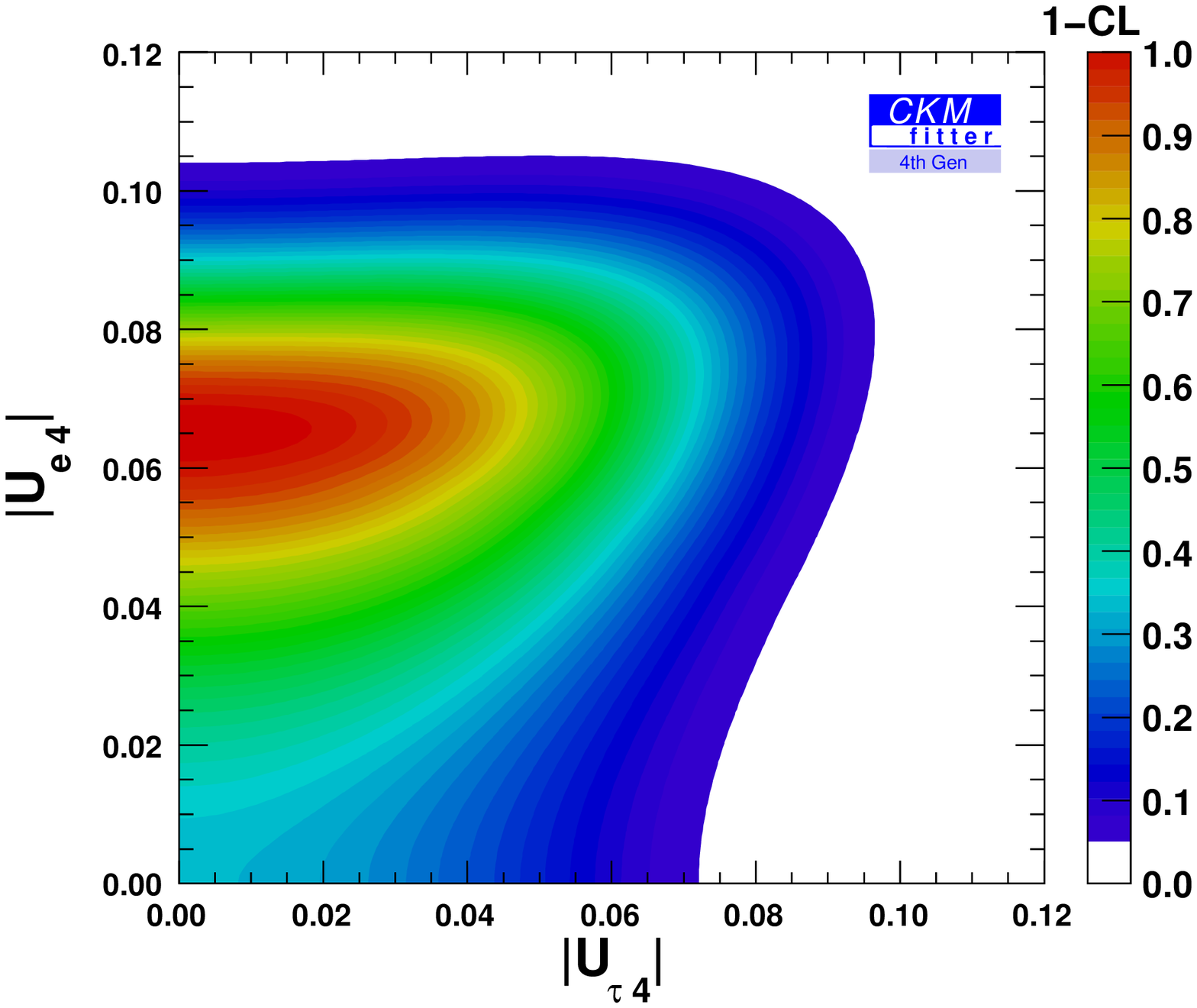}}
     \caption{Two-dimensional constraints on $|U_{e4}|$, $|U_{\mu 4}|$ and $|U_{\tau 4}|$. 
              Left column: results when leptonic $\mu$ and $\tau$ decays are combined. 
              Right column: results when radiative $\mu$ and $\tau$ decay
              measurements are taken into account as well. Please note the different
              ranges on both axis.  The color coding shows 1 - Confidence Level (1-CL).}
     \label{fig:Ue4Umu4Utau4}
\end{figure}
Figs.~\ref{fig:a_Ue4Umu4Utau4},~\ref{fig:c_Ue4Umu4Utau4}, and~\ref{fig:e_Ue4Umu4Utau4} 
show the actual constraints on $|U_{e 4}|$ versus $|U_{\mu 4}|$, $|U_{\tau 4}|$ versus 
$|U_{\mu 4}|$, and $|U_{e 4}|$ versus $|U_{\tau 4}|$ when using the input 
values $BF(\tau \rightarrow e \nu {\bar{\nu}})= 0.1785 \pm 0.0005$~\cite{PDG2008} 
and $BF(\tau \rightarrow \mu \nu {\bar{\nu}}) = 0.1736 \pm 0.0005$~\cite{PDG2008}
measured at LEP I,
$BF(\tau \rightarrow \mu \nu {\bar{\nu}})/BF(\tau \rightarrow e \nu {\bar{\nu}})=0.9796 \pm 0.0016 \pm 0.0036$
as recently measured by \babar~\cite{BABAR},
and the $\mu$ lifetime $\tau_{\mu}=(2.197034 \pm 0.000021) \times 10^{-6}~{\rm s}$~\cite{PDG2008}.
In the numerical analysis we explicitely take into account the correlation coefficient 
of $-13~\%$ between the LEP I measurements of $BF(\tau \rightarrow e \nu \bar{\nu})$ and 
$BF(\tau \rightarrow \mu \nu \bar{\nu})$ as quoted in~\cite{PDG2008}. We stress that
the uncertainties from $\tau_{\tau}$ and $m_{\tau}$ cancel in the constraint $|U_{e 4}|$ 
versus $|U_{\mu 4}|$, but not in $|U_{\tau 4}|$ versus $|U_{\mu 4}|$ and $|U_{e 4}|$ versus 
$|U_{\tau 4}|$.
One clearly sees the high correlation induced by the experimentally well-established 
lepton universality with the upper limit being close to the theoretical limit discussed 
just before. 
The observed correlation starts to vanish for values below about $0.1$ due to the finite 
experimental precision. In this region the constraint on $|U_{e 4}|$ versus $|U_{\mu 4}|$ 
(Fig.~\ref{fig:a_Ue4Umu4Utau4}) has a slightly asymmetric shape for small values of $|U_{e 4}|$ 
induced by the \babar measurement of 
$BF(\tau \rightarrow \mu \nu {\bar{\nu}})/BF(\tau \rightarrow e \nu {\bar{\nu}})$~\cite{BABAR}
which deviates by $1.8~\sigma$ from the expected value of 
$PS(m_{\mu},m_{\tau})/PS(m_{e},m_{\tau})=0.9726$ in the case of lepton universality.
In contrast, the LEP I measurements of $BF(\tau \rightarrow e \nu {\bar{\nu}})$ and 
$BF(\tau \rightarrow \mu \nu {\bar{\nu}})$ are in very good agreement with the 
expectation from lepton universality: 
$BF(\tau \rightarrow \mu \nu {\bar{\nu}})/BF(\tau \rightarrow e \nu {\bar{\nu}})=0.9725 \pm 0.0043$. 
The constraint $|U_{e 4}|$ versus $|U_{\tau 4}|$ shows an asymmetry at small values
as well (Fig.~\ref{fig:e_Ue4Umu4Utau4}) indicating a small but not significant deviation
from lepton universality between the electron and $\tau$ sector. In contrast $|U_{\tau 4}|$ 
versus $|U_{\mu 4}|$ (Fig.~\ref{fig:c_Ue4Umu4Utau4}) shows an almost perfect symmetry at 
small values reflecting at the current level of precision very good agreement of the 
data with the lepton universality hypothesis between the $\mu$ and the $\tau$ sector.

\section{The role of radiative decays $\mu \rightarrow e \gamma$,
         $\tau \rightarrow e \gamma$ and $\tau \rightarrow \mu \gamma$}
Observables which can provide constraints on products of PMNS matrix elements and hence 
introduce information which is ``orthogonal'' to the one obtained from leptonic decays 
are radiative $\mu$- and $\tau$-decay rates. We take the expression for the calculation 
of the radiative decay rates from Ref.~\cite{Altarelli}. In~\cite{Altarelli} the decay 
rate was calculated for a heavy third generation neutrino after the $\tau$ lepton had 
been discovered at SPEAR. To a very good approximation the result does not depend on the 
contributions from light neutrinos. By replacing the mixing parameters with corresponding 
PMNS matrix elements for the 4SM case one can write:
\begin{equation}\label{Eq:MuRadDecayRate}
\Gamma(\mu^{-} \rightarrow e^{-} \gamma)= \frac{G_{F}^{2} m_{\mu}^{5}}{192 \pi^{3}} \cdot 3 \cdot \frac{\alpha(m_{\mu})}{32 \pi} 
\frac{m_{\nu_{4}}^{4}}{m_{W}^{4}}
\cdot |U_{e 4}^{*} U_{\mu 4}|^{2} \left[ 6 (I^{(2)} \left( \frac{m_{\nu_{4}}^{2}}{m_{W}^{2}} \right) - I^{(3)} \left( \frac{m_{\nu_{4}}^{2}}{m_{W}^{2}} \right)) \right]^{2},
\end{equation} 
\begin{equation}\label{Eq:TauRadDecayRateE}
\Gamma(\tau^{-} \rightarrow e^{-} \gamma)= 
\frac{G_{F}^{2} m_{\tau}^{5}}{192 \pi^{3}} \cdot 3 \cdot \frac{\alpha(m_{\tau})}{32 \pi} \frac{m_{\nu_{4}}^{4}}{m_{W}^{4}}
\cdot |U_{e 4}^{*} U_{\tau 4}|^{2} 
\left[ 6 (I^{(2)} \left( \frac{m_{\nu_{4}}^{2}}{m_{W}^{2}} \right)-I^{(3)} \left( \frac{m_{\nu_{4}}^{2}}{m_{W}^{2}} \right)) \right]^{2},
\end{equation}
and
\begin{equation}\label{Eq:TauRadDecayRateMu}
\Gamma(\tau^{-} \rightarrow \mu^{-} \gamma)= 
\frac{G_{F}^{2} m_{\tau}^{5}}{192 \pi^{3}} \cdot 3 \cdot \frac{\alpha(m_{\tau})}{32 \pi} \frac{m_{\nu_{4}}^{4}}{m_{W}^{4}}
\cdot |U_{\mu 4}^{*} U_{\tau 4}|^{2} 
\left[ 6 (I^{(2)} \left( \frac{m_{\nu_{4}}^{2}}{m_{W}^{2}} \right)-I^{(3)} \left( \frac{m_{\nu_{4}}^{2}}{m_{W}^{2}} \right) ) \right]^{2}
\end{equation}
with $m_{\nu_{4}}$ being the mass of the heavy fourth generation neutrino
and 
\begin{equation}\label{Eq:InamiLim}
I^{(n)}(x)= \int_{0}^{1} \frac{dz \cdot z^{n}}{z + x \cdot (1-z)}.
\end{equation}
It is important to note that the decay rate for these radiative decays
do not vanish in the limit $m_{\nu_{4}} \rightarrow \infty$ since
the neutrino Yukawa-coupling increases with increasing $m_{\nu_{4}}$
which compensates the decrease of the loop integral. Therefore one
can not escape the constraints from radiative lepton decays.

Since the mass limits from LEP2 on a fourth generation neutrino are only 
valid for unstable neutrinos~\cite{LEPLimitChargedLepton} these limits 
depend implicitely on the mass of the heavy charged lepton and on the 
size of the PMNS matrix elements. To extract numerical results that are 
independent from these assumptions we choose for the heavy neutrino mass 
the robust constraint $m_{\nu_4} > m_{Z}/2$ resulting from the LEP1 
measurements of the $Z$-resonance~\cite{LEPZwidth}. This is a conservative
choice since for higher masses the constraints are getting stronger.
For $m_{\nu_4} = 45~{\rm GeV}$ one finds 
$\left[ 6 (I^{(2)} \left( \frac{m_{\nu_{4}}^{2}}{m_{W}^{2}} \right)-I^{(3)} \left( \frac{m_{\nu_{4}}^{2}}{m_{W}^{2}} \right) ) \right]^{2} \approx 0.513$.

\begin{figure}
     \centering
     \subfigure[]{
          \label{fig:a_LepRadTauMuKl3Pil2}
          \includegraphics[width=.475\textwidth]{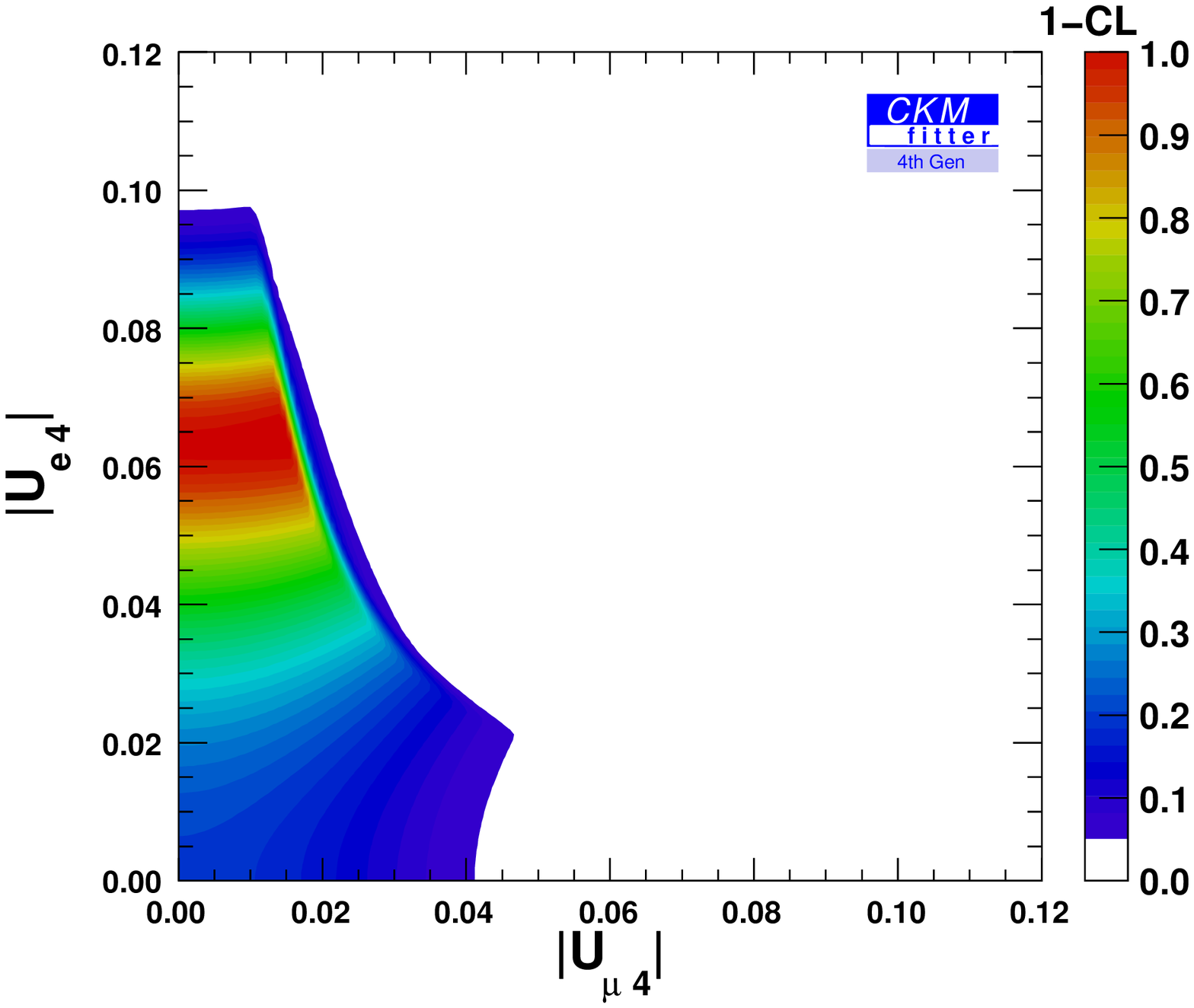}}
     \hspace{.01in}
     \subfigure[]{
          \label{fig:b_LepRadTauMuKl3Pil2}
          \includegraphics[width=.475\textwidth]{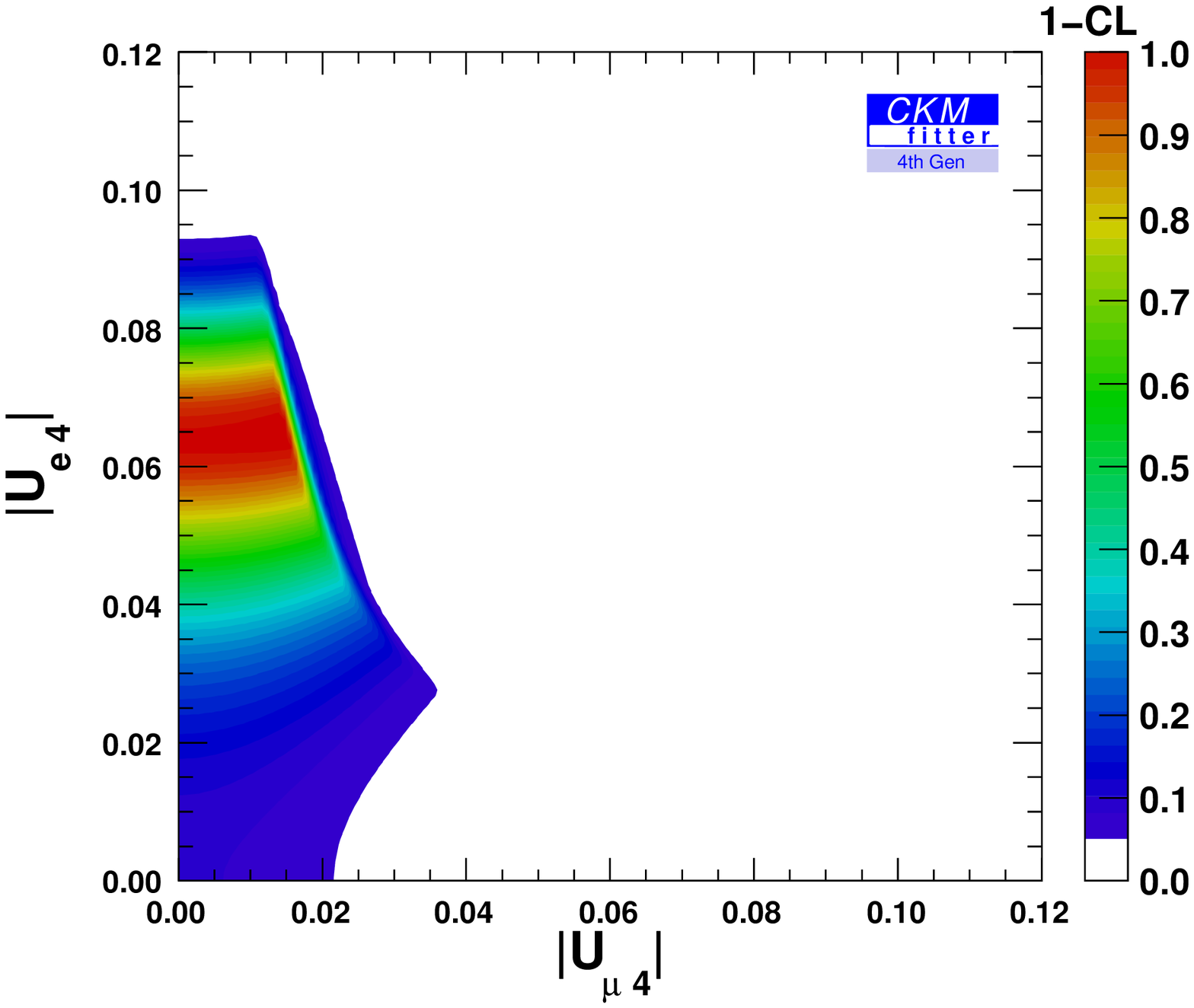}}\\
     \vspace{.01in}
     \subfigure[]{
          \label{fig:c_LepRadTauMuKl3Pil2}
          \includegraphics[width=.475\textwidth]{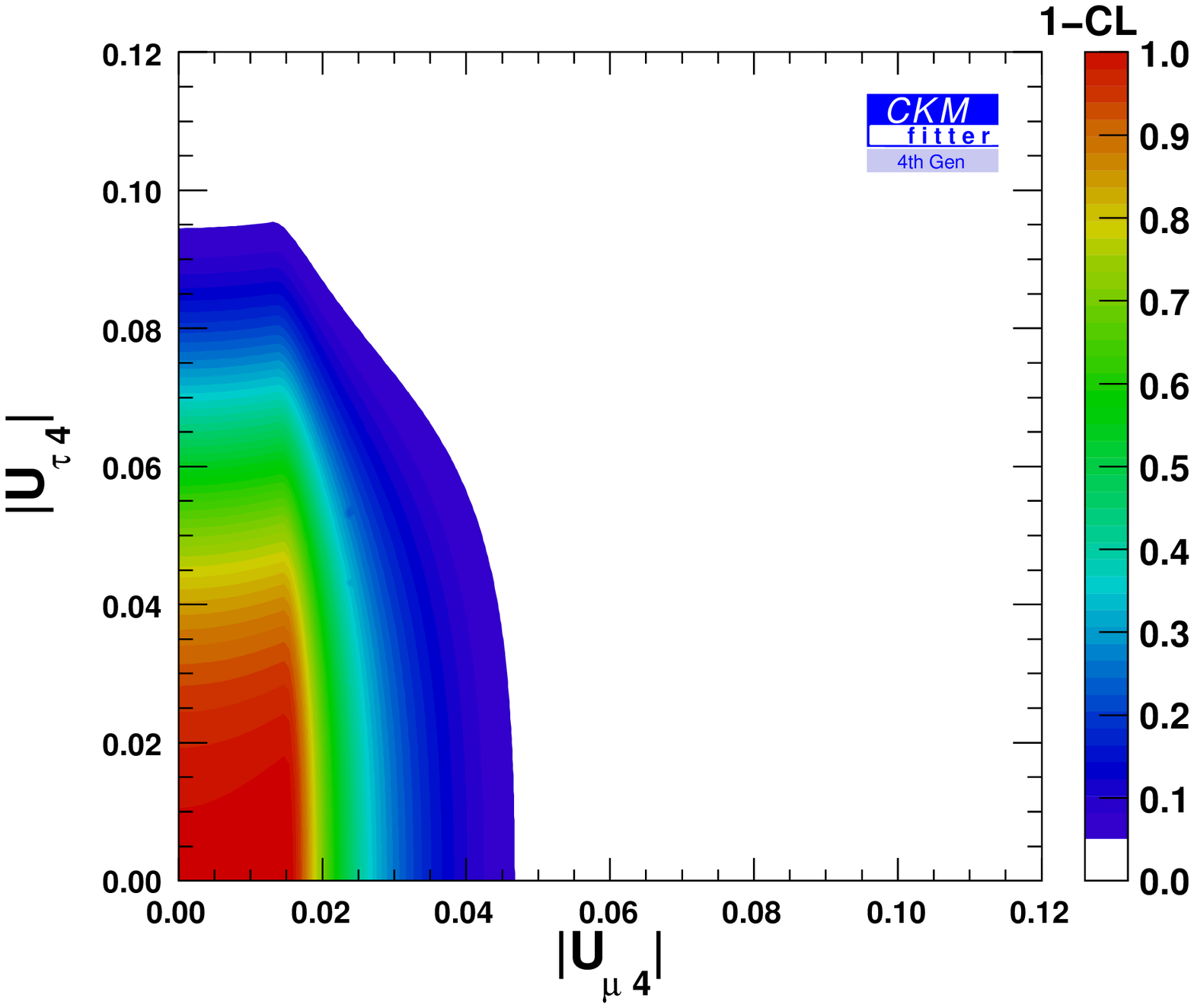}}
     \hspace{.01in}
     \subfigure[]{
          \label{fig:d_LepRadTauMuKl3Pil2}
          \includegraphics[width=.475\textwidth]{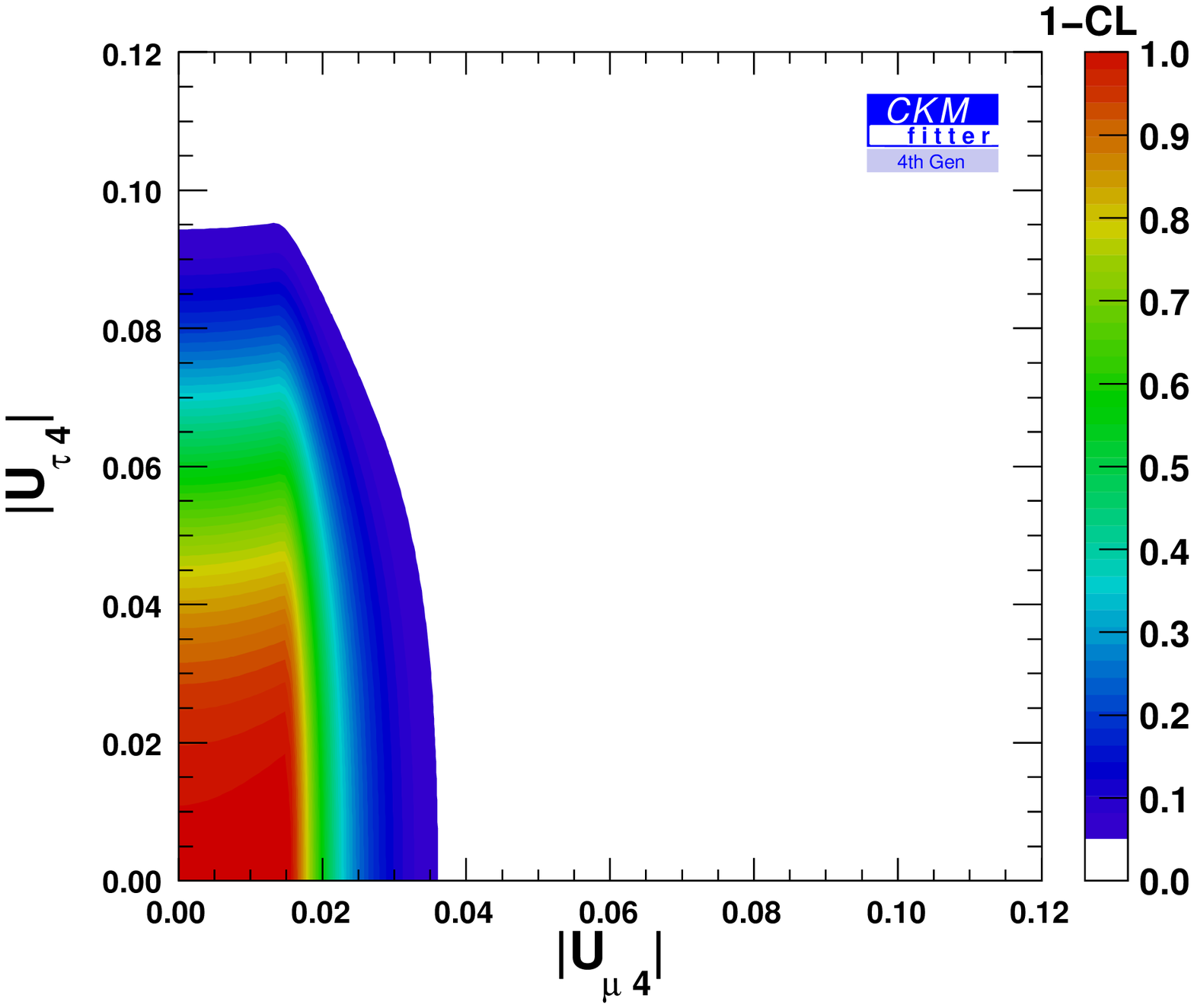}}\\
     \vspace{.01in}
     \subfigure[]{
          \label{fig:e_LepRadTauMuKl3Pil2}
          \includegraphics[width=.475\textwidth]{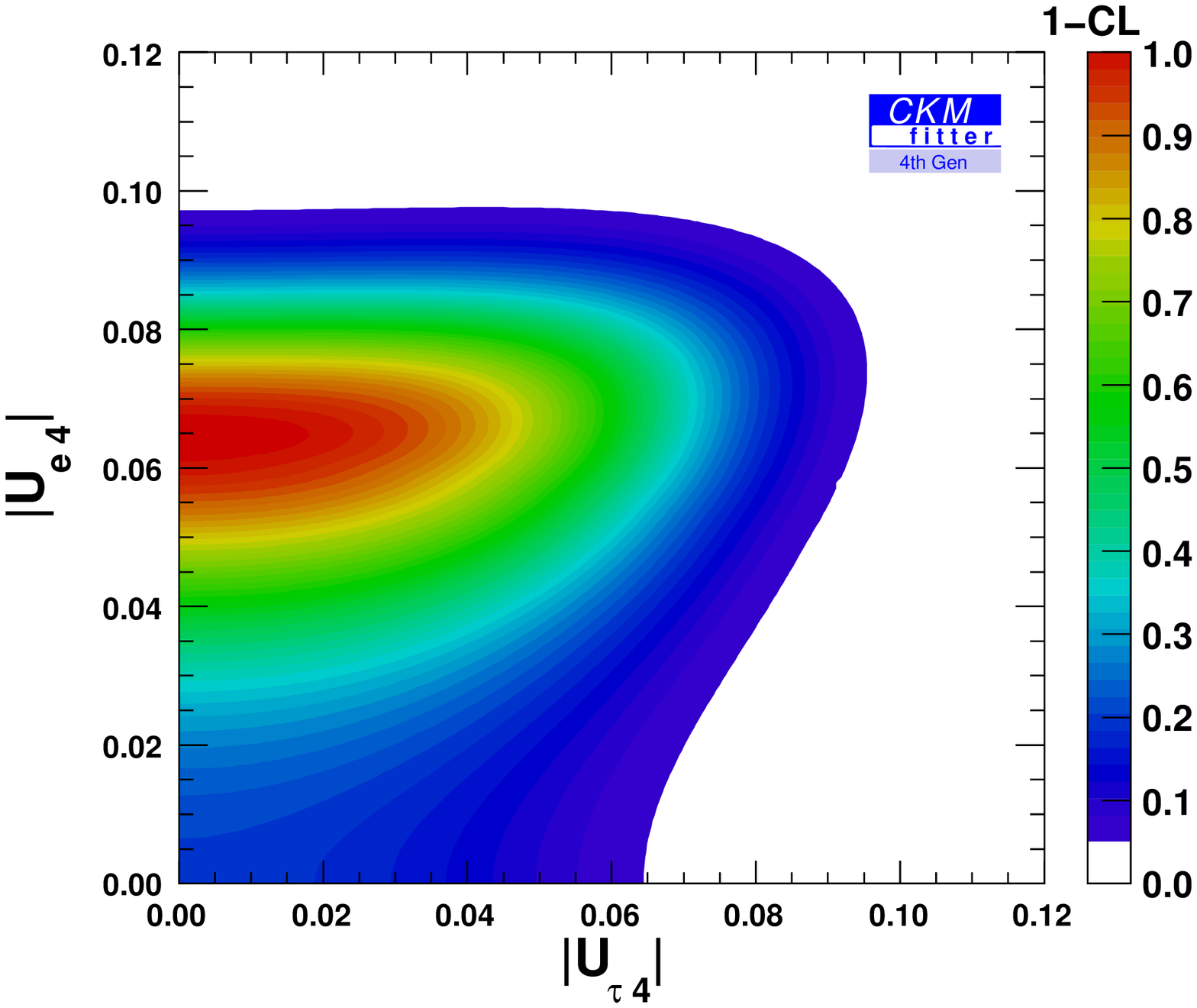}}
     \hspace{.01in}
     \subfigure[]{
          \label{fig:f_LepRadTauMuKl3Pil2}
          \includegraphics[width=.475\textwidth]{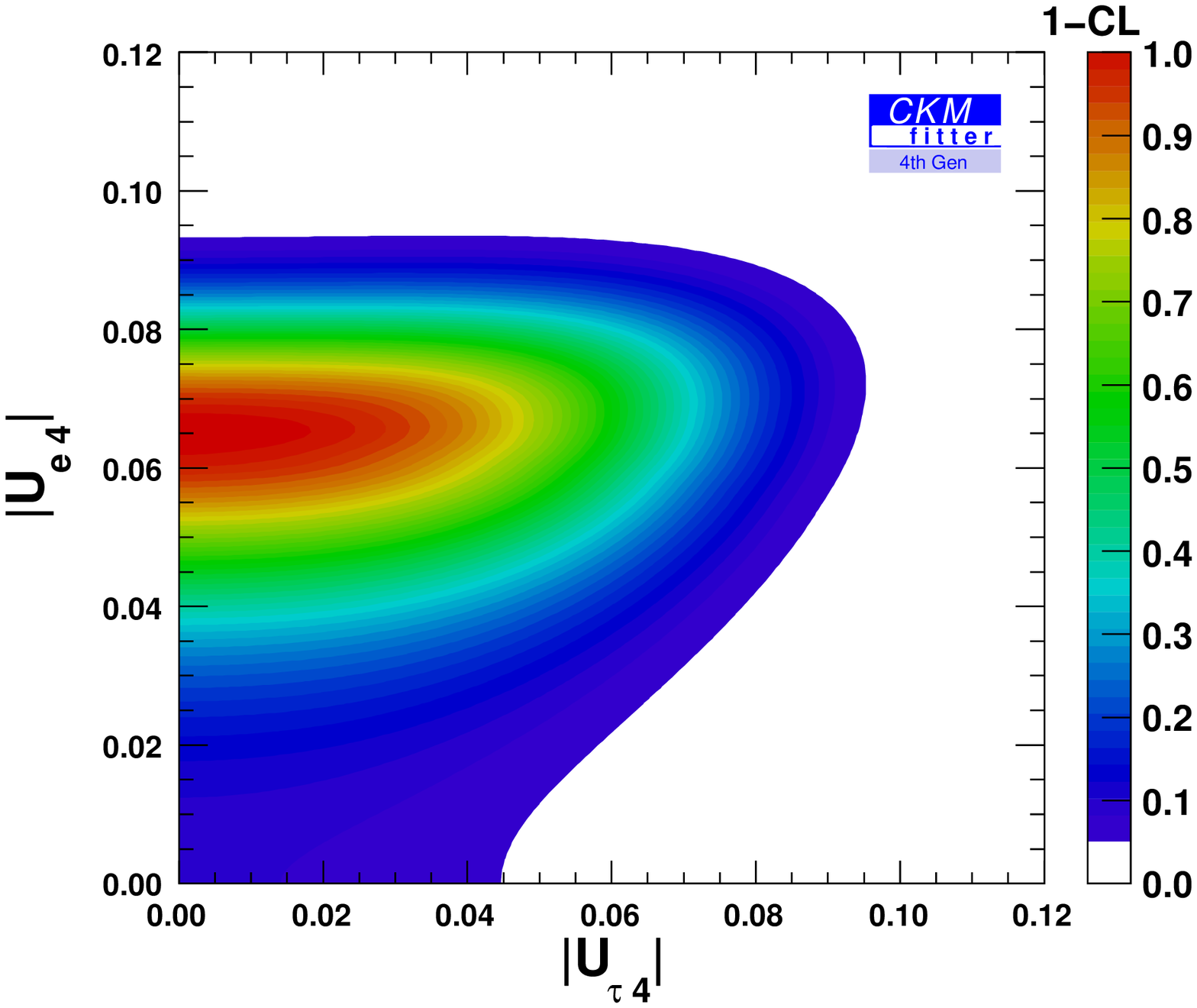}}
     \caption{Two-dimensional constraints on $|U_{e4}|$, $|U_{\mu 4}|$ and $|U_{\tau 4}|$. 
              Left column: results when leptonic and radiative $\mu$, respectively, $\tau$ 
              decays are combined with $K_{\ell 3}$ measurements. Right column: results 
              when $\pi_{\ell 2}$ measurements are taken into account as well.}
     \label{fig:LepRadTauMuKl3Pil2}
\end{figure}

We use the following experimental search limits: 
$BF(\mu \rightarrow e \gamma) < 1.2 \times 10^{-11}$~\cite{LimitMuEGamma},
$BF(\tau \rightarrow e \gamma) < 3.3 \times 10^{-8}$~\cite{LimitTauRadiativeBabar}, 
and $\tau \rightarrow \mu \gamma < 4.4 \times 10^{-8}$~\cite{LimitTauRadiativeBabar}
where all limits are quoted at $90~\%$ CL.
Figs.~\ref{fig:b_Ue4Umu4Utau4},\ref{fig:d_Ue4Umu4Utau4}, and \ref{fig:f_Ue4Umu4Utau4} 
show the constraints on $|U_{e 4}|$ versus $|U_{\mu 4}|$, $|U_{e 4}|$ versus 
$|U_{\tau 4}|$, and $|U_{\tau 4}|$ versus $|U_{\mu 4}|$ obtained by combining 
the leptonic $\mu$ and $\tau$ decays discussed in the previous section together 
with the search limits for the radiative decays $\mu \rightarrow e \gamma$, 
$\tau \rightarrow e \gamma$ and $\tau \rightarrow \mu \gamma$.
The constraint from $BF(\mu \rightarrow e \gamma)$ imposes a constraint of 
the form $|U_{e 4}| = const./|U_{\mu 4}|$ corresponding to a hyperbolic shape.
When combined with the constraint depicted in Fig.~\ref{fig:a_Ue4Umu4Utau4} this
hyperbolic shape becomes clearly visible in Fig.~\ref{fig:b_Ue4Umu4Utau4}.
Due to the much weaker search limits for the radiative $\tau$ decays this particular 
shape is barely observable in Fig.~\ref{fig:d_Ue4Umu4Utau4} and not visible in
Fig.~\ref{fig:f_Ue4Umu4Utau4}.  
These plots demonstrate that the constraints from leptonic $\mu$ and $\tau$ decays 
when combined with the search limits from the radiative $\mu$ and $\tau$ decays 
are narrowed with respect to the leptonic $\mu$ and $\tau$ constraints only. 
The constraints lead to the following PMNS matrix:\\
              \begin{equation}
              \label{Eq:PMNSmatrixLeptRad}
              U_{PMNS} = 
              \left(
              \begin{array}{cccc}
               *      & *      & *      & <0.099 \\
               *      & *      & *      & <0.048 \\
               *      & *      & *      & <0.086 \\
               <0.122 & <0.122 & <0.122 & >0.9925
              \end{array}
              \right)
              \end{equation}\\
where we quote $2\sigma$ limits and do not quantify any constraints
on the $3 \times 3$ submatrix for the first three light neutrino 
generations which is determined by oscillation experiments.\\
As a result of the improved constraints on the fourth generation PMNS matrix
elements, the precision of $G_F$ is considerably improved in the combined fit 
as well: $G_F^{3SM} \le G_F^{4SM} \le 1.0050 \cdot G_F^{3SM}$ ($2\sigma$ level). 
That is, $G_F$ could still be up to $0.50~\%$ larger than the value obtained 
when assuming $3 \times 3$ unitarity for the first three generations in the 
neutrino sector.\\
Figs.~\ref{fig:b_Ue4Umu4Utau4},~\ref{fig:d_Ue4Umu4Utau4}, and~\ref{fig:f_Ue4Umu4Utau4} 
also reveal that the precision on the fourth generation PMNS matrix elements 
is limited by how well lepton universality is tested by the measurement of 
the $\mu$ lifetime and the leptonic $\tau$ decays. 

\section{Leptonic and semileptonic decays of kaons and pions}
In addition to leptonic $\mu$ and $\tau$ decays one can test lepton universality
for example in leptonic and semileptonic kaon decays.\\
In the ratio of branching fractions 
$R_{K}=BF(K^{+} \rightarrow e^{+} \nu_{e})/BF(K^{+} \rightarrow \mu^{+} \nu_{\mu})$
the dependence on the CKM element ${|V_{us}|}^{2}$ as well as on $G_{F}$ 
and other factors like the decay constant cancels and one can constrain 
$(1-|U_{e 4}|^{2})/(1-|U_{\mu 4}|^{2})$.
The most precise value for $R_{K}$ has been presented recently by the NA62 
collaboration as a preliminary result with a relative uncertainty of 
$0.7~\%$~\cite{NA62Talk}. This result has still a lower precision than the 
ratio of branching fractions between the corresponding leptonic $\tau$ 
decays (relative uncertainty of around $2.9~\%$ when the LEP I and \babar
results are combined). When taking into account the new NA62 result the world
average $R_{K}=(2.498 \pm 0.014)~\%$~\cite{NA62Talk} has a relative uncertainty
of $0.56~\%$ to be compared to the 3SM prediction of 
$R_{K}^{theo}=(2.477 \pm 0.001)~\%$~\cite{CiriglianoRosell}.
Since the NA62 result is still to be considered preliminary and the world 
average for $R_{K}$ without this result has a quite large uncertainty of 
$0.97~\%$ we do not use this input in our fit. The final relative uncertainty 
on $R_{K}$ anticipated by NA62 is $0.4~\%$ which should have a significant 
impact on lepton universality tests.\\
In contrast to leptonic kaon decays, semileptonic $K_{e3}$ and $K_{\mu 3}$ 
branching fractions provide already a test of lepton universality 
competitive with leptonic $\mu$ and $\tau$ decays~\cite{flavianet}.
The semileptonic kaon decay rates including radiative corrections are taken 
from Ref.~\cite{Kl3DecayRate} and modified for the fourth generation scenario 
as
\begin{equation}\label{Eq:Kl3DecayRate}
\Gamma(K_{\ell 3 (\gamma)})= 
\frac{G_{F}^{2} m_{K}^{5}}{192 \pi^{3}} C_{K} S_{ew} f_{+}(0)^{2} |V_{us}|^{2} I_{K}^{\ell}(\lambda_{+,0})
(1 + \delta_{SU(2)}^{K} + \delta_{em}^{K \ell})^{2}
\cdot (1-|U_{\ell 4}|^{2}),
\end{equation} 
where $\ell=e$ or $\mu$, $V_{us}$ is the CKM matrix element for the weak transition
between a s- and u-quark, and $C_{K}$ is a Clebsch-Gordan coefficient being equal 
to 1 ($1/\sqrt{2}$) for a neutral kaon in the initial state leading to a charged 
pion in the final state (for a charged kaon in the initial state leading to a 
neutral pion in the final state). $I_{K}^{\ell}(\lambda_{+,0})$ is a form-factor 
dependent phase integral where $\lambda_{+}$ and $\lambda_{0}$ are the form factor 
slope parameters for the charged, respectively, neutral
kaon decay. $S_{ew}$ is the short-distance electroweak correction calculated 
in Ref.~\cite{Sirlin2}. $\delta_{SU(2)}^{K}$ is a long-distance isospin-breaking 
correction, and $\delta_{em}^{K \ell}$ takes into account the long-distance QED 
corrections~\cite{cirigliano}. 
The vector form-factor at zero-momentum transfer is denoted by $f_{+}(0)$.\\
By comparing $K_{\ell 3 (\gamma)}$ decay rates for electron and muon final 
states for a charged, respectively, a neutral kaon in the initial state most 
of the factors cancel. The Flavianet Kaon Working Group provides results for 
the quantities $|V_{us}| f_{+}(0)$ separated between charged and neutral kaon 
initial states as well as for leptonic final states, and it is also possible 
to extract the uncorrelated and correlated uncertainties from the information 
provided in~\cite{flavianet}.
In the fourth generation scenario these quantities have to be reinterpreted 
as $|V_{us}| f_{+}(0) \sqrt{(1-|U_{\ell 4}|^{2})} G_{F}^{4SM}/G_{F}^{3SM}$.
With these additional inputs we obtain tightened constraints on $|U_{e 4}|$, 
$|U_{\mu 4}|$, and $|U_{\tau 4}|$ as shown in Figs.~\ref{fig:a_LepRadTauMuKl3Pil2},
~\ref{fig:c_LepRadTauMuKl3Pil2}, and~\ref{fig:e_LepRadTauMuKl3Pil2}.
One notes that the constraint on $|U_{\mu 4}|$ becomes significantly tighter
than on $|U_{e 4}|$. This is understood as follows: the extracted 
$|V_{us}| f_{+}(0)$ value for the $K_{e3}$ decays is smaller than for 
$K_{\mu 3}$, although they are still in agreement at the $1\sigma$ level, 
which leads to a higher preferred value of $|U_{e 4}|$ compared to $|U_{\mu 4}|$.
The PMNS matrix has then the following $2\sigma$ limits:\\
              \begin{equation}
              \label{Eq:PMNSmatrixLeptRadKl3}
              U_{PMNS} = 
              \left(
              \begin{array}{cccc}
               *      & *      & *      & <0.092 \\
               *      & *      & *      & <0.037 \\
               *      & *      & *      & <0.085 \\
               <0.117 & <0.117 & <0.117 & >0.9932
              \end{array}
              \right).
              \end{equation}\\
The allowed range for $G_{F}$ is further tightened: 
$G_F^{3SM} \le G_F^{4SM} \le 1.0044 \cdot G_F^{3SM}$ ($2\sigma$ level). 

\begin{figure}
     \centering
     \subfigure[]{
          \label{fig:a_GFLepRadTauMuKl3Pil2}
          \includegraphics[width=.475\textwidth]{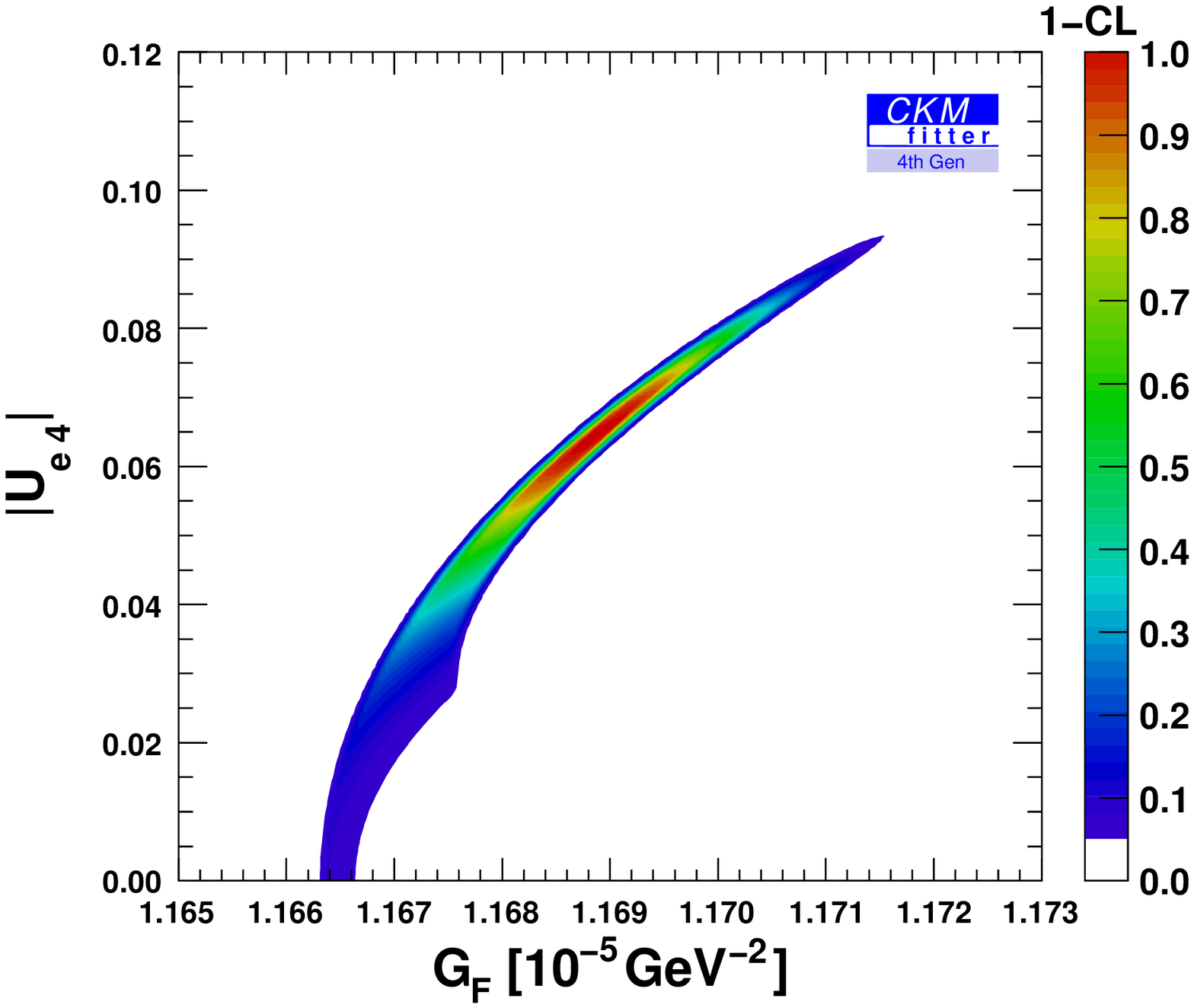}}
     \hspace{.01in}
     \subfigure[]{
          \label{fig:b_GFLepRadTauMuKl3Pil2}
          \includegraphics[width=.475\textwidth]{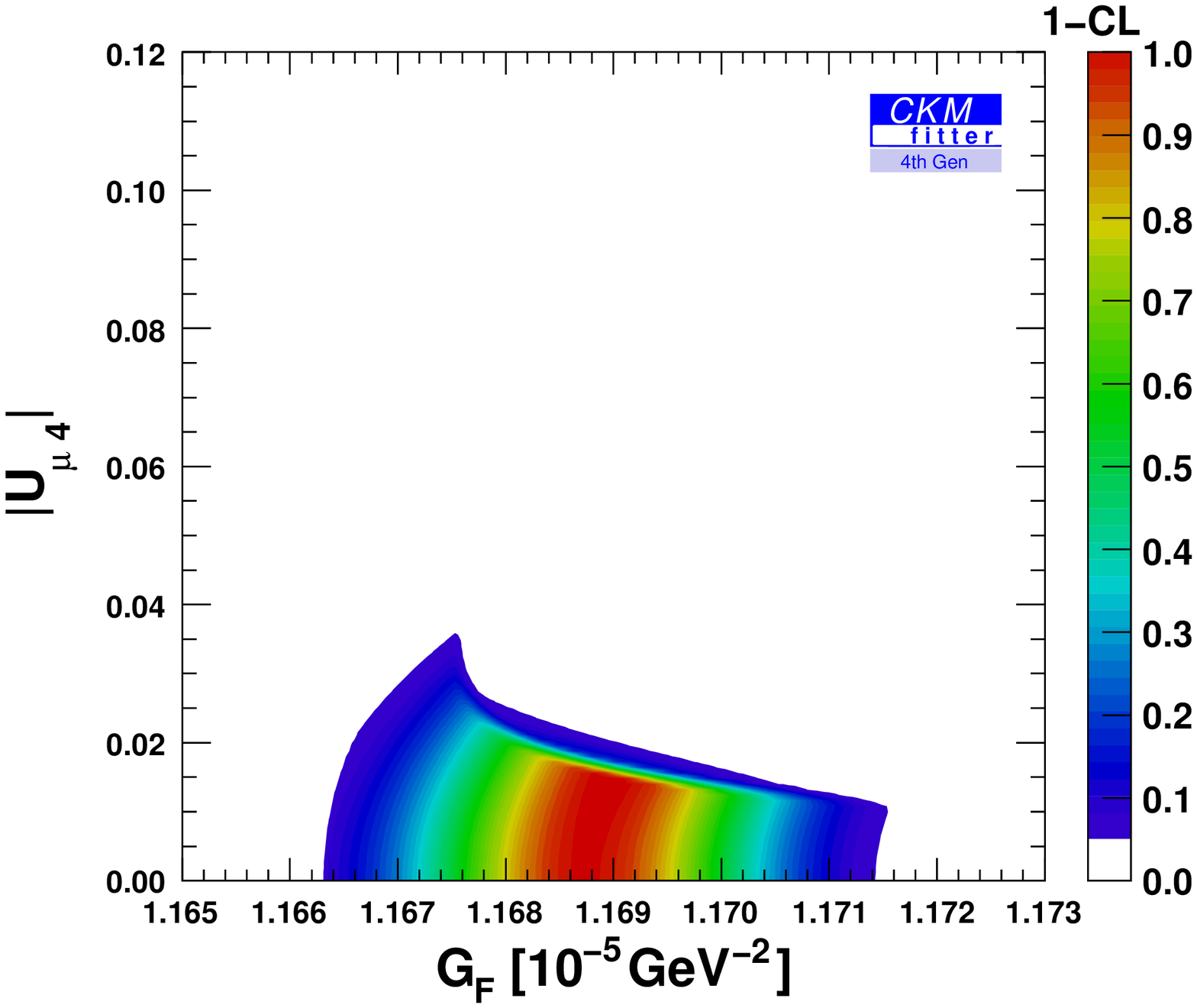}}\\
     \vspace{.01in}
     \subfigure[]{
          \label{fig:c_GFLepRadTauMuKl3Pil2}
          \includegraphics[width=.475\textwidth]{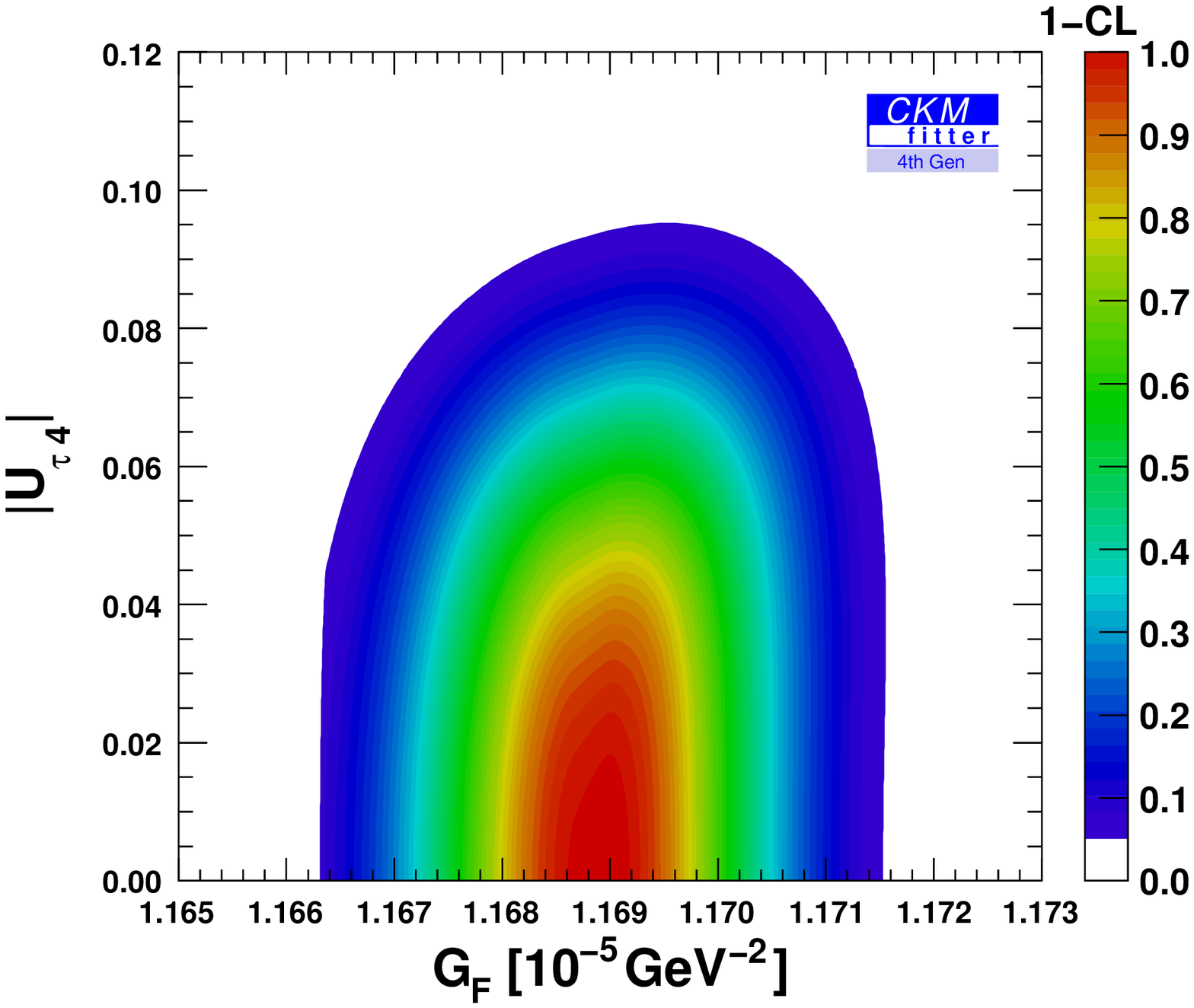}}
     \hspace{.01in}
     \subfigure[]{
          \label{fig:d_GFLepRadTauMuKl3Pil2}
          \includegraphics[width=.475\textwidth]{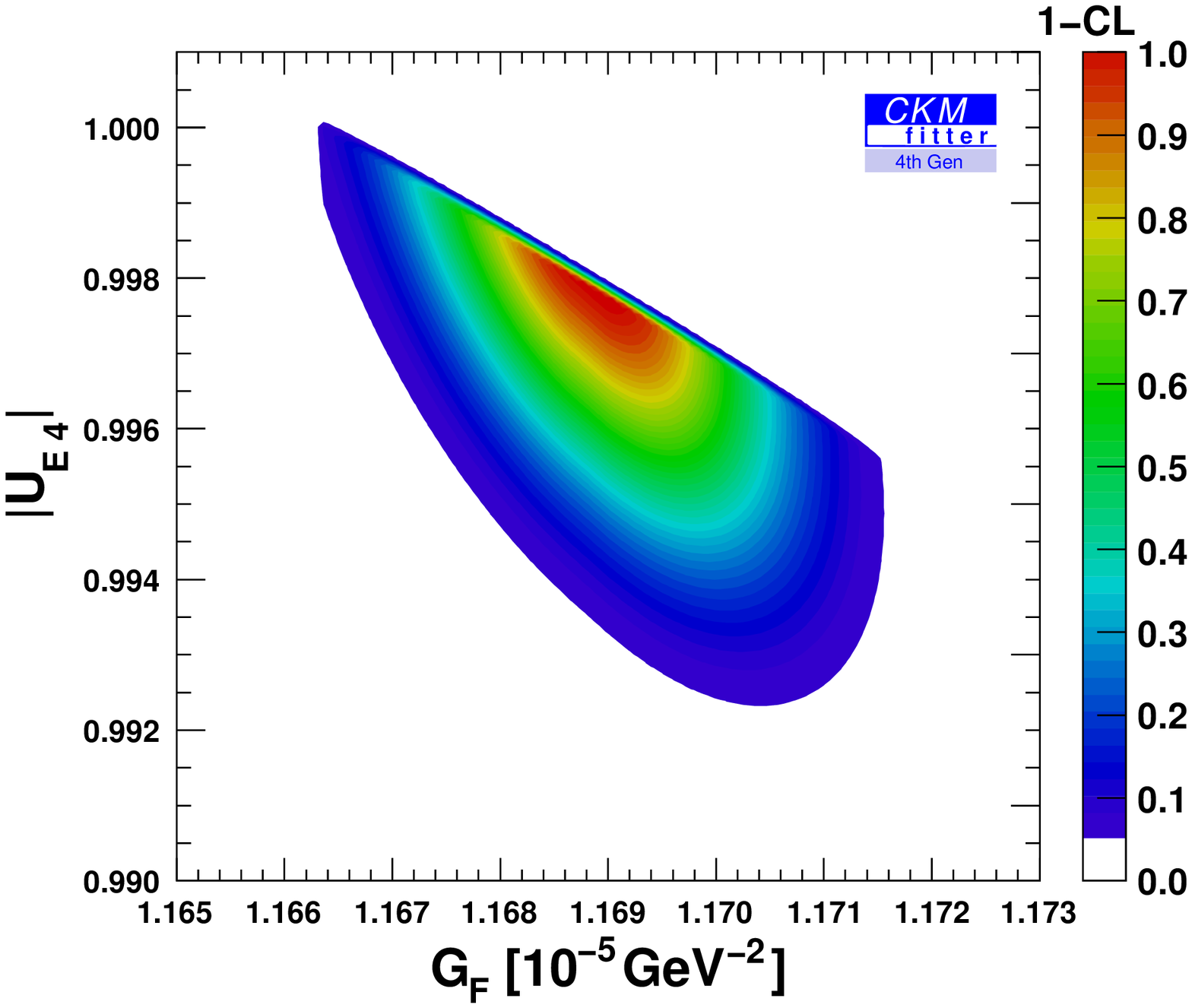}}\\
     \caption{Two-dimensional constraints on $|U_{e4}|$, $|U_{\mu 4}|$, $|U_{\tau 4}|$
              and $|U_{E 4}|$ as a function of $G_F$.}
     \label{fig:GFLepRadTauMuKl3Pil2}
\end{figure}

Analogous to leptonic Kaon decays the ratio of branching fractions for charged pions
to leptonic final states ($\pi_{\ell 2}$ decays)
\begin{equation}
R_{\pi} = \frac{BF(\pi^{+} \rightarrow e^{+} \nu_{e})}{BF(\pi^{+} \rightarrow \mu^{+} \nu_{e})}
\end{equation}
allows to constrain $(1-|U_{e 4}|^{2})/(1-|U_{\mu 4}|^{2})$. Using the experimental 
measurements 
$BF(\pi^{+} \rightarrow e^{+} \nu_{e}(\gamma))=(1.230 \pm 0.004) \cdot 10^{-4}$~\cite{PDG2008}
(which is and average of the measurements quoted in Refs.~\cite{Czapek,Britton,Bryman}) 
and $BF(\pi^{+} \rightarrow \mu^{+} \nu_{\mu}(\gamma)=0.9998770 \pm 0.0000004$~\cite{PDG2008}, 
and the theoretical prediction 
$R_{\pi}^{theo}=(1.2354 \pm 0.0002) \cdot 10^{-4}$~\cite{Finkemeier} we obtain 
the constraints on $|U_{e 4}|$, $|U_{\mu 4}|$, and $|U_{\tau 4}|$ as shown in 
Figs.~\ref{fig:b_LepRadTauMuKl3Pil2},~\ref{fig:d_LepRadTauMuKl3Pil2}, 
and~\ref{fig:f_LepRadTauMuKl3Pil2}.\\
For the PMNS matrix the $2\sigma$ limits read then:
              \begin{equation}
              \label{Eq:PMNSmatrixLeptRadKl3Pil2}
              U_{PMNS} = 
              \left(
              \begin{array}{cccc}
               *      & *      & *      & _{>0.021}^{<0.089} \\
               *      & *      & *      & <0.029 \\
               *      & *      & *      & <0.085 \\
               <0.115 & <0.115 & <0.115 & _{>0.9934}^{<0.9998}
              \end{array}
              \right)
              \end{equation}
One should note the non-trivial lower limit for $|U_{e 4}|$ and the non-trivial upper limit 
for $|U_{E 4}|$ where we denote the heavy charged lepton as $E$. The p-value for the 3SM 
point $|U_{e 4}|=0$ is now $2.6~\%$. The allowed range for $G_{F}$ is again tightened: 
$1.0002 \cdot G_F^{3SM} \le G_F^{4SM} \le 1.0040 \cdot G_F^{3SM}$ ($2\sigma$ level).
The lower limit on $G_F^{4SM}$ being larger than the $G_F^{3SM}$ value is directly related 
to the lower (upper) limit on $|U_{e 4}|$ ($|U_{E 4}|$). These correlations are illustrated
in Fig.~\ref{fig:a_GFLepRadTauMuKl3Pil2}-\ref{fig:d_GFLepRadTauMuKl3Pil2} where we show the 
two-dimensional constraints on $|U_{e 4}|$, $|U_{\mu 4}|$, $|U_{\tau 4}|$ and $|U_{E 4}|$ 
as a function of $G_{F}$.

\section{Discussion}
We discuss first the consequences related to the loss in precision in the extracted $G_{F}$ 
value and the dependence of traditional CKM observables on PMNS matrix elements.
\begin{itemize}
  \item CKM elements extracted from leptonic and semileptonic meson decays have to be 
        separated according to the leptonic final state.
  \item Within a four generation scenario the extraction of $|V_{ud}|$ from superallowed 
        $\beta$ decays is significantly affected in its precision: one extracts 
        $|V_{ud}| \sqrt{(1-|U_{e 4}|^{2})} G_F^{4SM}/G_F^{3SM}$ instead of $|V_{ud}|$. 
        In a recent analysis, Hardy and Towner find $|V_{ud}|=0.97425 \pm 0.00022$~\cite{TownerHardy}.
        Interpreting this number as $|V_{ud}| \sqrt{(1-|U_{e 4}|^{2})} G_F^{4SM}/G_F^{3SM}$
        and performing a combined fit with all our inputs to constrain $G_F^{4SM}$ and the 
        four generation PMNS elements results in $|V_{ud}|=0.97413^{+0.00033}_{-0.00023}$ at 
        $1\sigma$ level, $|V_{ud}|=0.97413^{+0.00053}_{-0.00046}$ at $2\sigma$ level and
        $|V_{ud}|=0.97425^{+0.00071}_{-0.00100}$ at $3\sigma$ level. As a consequence, the 
        precision in the extracted $|V_{ud}|$ value is significantly lower compared to the 
        3SM case.
        This reduction in precision for $|V_{ud}|$ will impact the constraints on fourth 
        generation CKM elements imposed by $4 \times 4$ unitarity. Other constraints on 
        CKM matrix elements are barely affected at the current level of precision.
  \item Ratios of CKM matrix elements determined from leptonic and semileptonic meson 
        decays can be extracted without any change as long as measured rates are used 
        with same final state leptons.\\ 
        Also the ratio $R=|V_{tb}|^{2}/(|V_{td}|^{2}+|V_{ts}|^{2}+|V_{tb}|^{2})$ 
        extracted from $p \bar{p} \rightarrow t \bar{t} X$ events with zero, one or two 
        b-tagged jets is unaffected.
  \item The CKM matrix elements $|V_{cd}|$ and $|V_{cs}|$ when extracted from deep inelastic 
        scattering of $\nu_{\mu}$ and ${\bar{\nu}}_{\mu}$ on nucleons can be determined
        without modifications. In this case, one measures the ratio of yields between dimuon 
        and single muon events~\cite{vcdCdhs,vcdCcfr,vcdCharmii}.
        The muon common to both event classes originates from the
        charged current transition between the muon neutrino and the muon.
        As a consequence, the dependence on $1-|U_{\mu 4}|^2$ cancels in the ratio
        and one extracts  $|V_{cd}|^2 \times B_{c} = (4.63 \pm 0.34) \times 10^{-3}$, 
        where $B_{c} = 0.0919 \pm 0.0094$~\cite{bolton,ushida,kubota} is the measured
        branching fraction of semileptonic decays of charmed hadrons.
  \item Within the 4SM scenario the reduced precision in $G_F$ requires a new 
        study of the electroweak precision fit since the usual strategy in these fits is 
        to neglect the uncertainty coming from $G_F$. In such a fit the dependence of the 
        electroweak precision observables on PMNS and CKM matrix elements in the self-energy 
        terms needs to be taken into account. Observables like the $W$-boson mass or the 
        partial rate $\Gamma(Z \rightarrow l^{+} l^{-})$ depend at leading order on $G_F$. 
        Only at higher order do the fourth generation particles enter with a corresponding 
        dependence on CKM and PMNS matrix elements. In turn, a consistent fourth generation 
        electroweak precision fit might strengthen the constraint on $G_F$.
\end{itemize}
Next, we discuss the bounds found for the fourth generation PMNS matrix elements in more detail.
\begin{itemize}
  \item In the past there have been studies on fourth generation PMNS matrix elements using 
        leptonic $\tau$ decays at a time when there was a discrepancy between the corresponding 
        branching fraction values, the $\tau$ lifetime and the $\tau$ mass~\cite{SwainTaylor}. 
        Only after an improved determination of the $\tau$ mass leading to a significant
        shift in the measured value made the discrepancy vanish.\\
        In the analysis of Ref.~\cite{SwainTaylor} it was assumed that there is only mixing 
        between the third and fourth generation. Hence, the determination of $G_F$ from the 
        muon lifetime was considered to be unaffected by the presence of a fourth generation.\\
        As a consequence, our analysis is much more general: it is able to extract limits on all
        PMNS matrix elements related to the fourth generation and to determine simultaneously 
        the Fermi constant $G_F$.
  \item In Ref.~\cite{Plehn} the radiative decay $\mu \rightarrow e \gamma$ was used, assuming 
        $G_{F}=G_{F}^{3SM}$, to set a limit on $|U_{e4} U_{\mu 4}^{*}|$. 
        However, no limit on the individual fourth generation PMNS elements could be set in 
        Ref.~\cite{Plehn} since the leptonic $\mu$ and $\tau$ decays have not been considered. 
        To obtain the limit of $|U_{e4} U_{\mu 4}^{*}|<4 \cdot 10^{-4}$ the authors of 
        Ref.~\cite{Plehn} used the search limit for the heavy neutrino mass, 
        $m_{\nu_{4}}=90.3~{\rm GeV}$\cite{LEPLimitChargedLepton}. 
        Within our analysis the limit on this product is significantly looser,
        $|U_{e4} U_{\mu 4}^{*}| < 1.16 \cdot 10^{-3}$ ($2\sigma$ level), since we conservatively 
        take $m_{\nu_{4}}>m_{Z}/2$. It has to be stressed though that our individual limits on 
        $|U_{e 4}|$, $|U_{\mu 4}|$, and $|U_{\tau 4}|$ would not change very much if 
        $m_{\nu_{4}}>90.3~{\rm GeV}$ were used in our analysis. 
  \item Recently, \babar has presented measurements of $BF(\tau \rightarrow h \nu_{\tau})$ 
        ($h=\pi, K$)~\cite{BABAR}. When compared to $BF(h \rightarrow \mu \nu_{\mu})$ \babar 
        finds a $3\sigma$ deviation from lepton universality in the $\mu-\tau$ sector. 
        Interpreted in a four generation scenario the \babar result~\cite{BABAR} reads 
        $\sqrt{1-|U_{\tau 4}|^{2}}/\sqrt{1-|U_{\mu 4}|^{2}}=0.985 \pm 0.005$. In contrast,
        we find $\sqrt{1-|U_{\tau 4}|^{2}}/\sqrt{1-|U_{\mu 4}|^{2}}=1.0001^{+0.0001}_{-0.0017}$ 
        ($1\sigma$ level) in our analysis which does not use the $\tau \rightarrow h \nu_{\tau}$ 
        data. As a consequence, we do not observe such a deviation from lepton universality.
  \item Leptonic $W$ decays $W \rightarrow \ell \nu_{\ell}$ ($\ell = e, \mu, \tau$) can 
        be used as well to test lepton universality. However, the experimental precision 
        of the measured branching fractions 
        $BF(W \rightarrow e \nu_{e})= 0.1075 \pm 0.0013$~\cite{PDG2008}, 
        $BF(W \rightarrow \mu \nu_{\mu})= 0.1057 \pm 0.0015$~\cite{PDG2008}, 
        $BF(W \rightarrow \tau \nu_{\tau})= 0.1125 \pm 0.0020$~\cite{PDG2008}, 
        can not compete with the other lepton universality tests studied in our analysis.
        It is important to note though that for leptonic $W$ decays the preferred but not 
        significant hierarchy is $|U_{\tau 4}| < |U_{e 4}| < |U_{\mu 4}|$.
        As a consequence, including the leptonic $W$ branching fractions does not change 
        significantly the constraints due to the lower precision but leads to an 
        increase of the $\chi^{2}$ value of the combined fit from $4.1$ (without $W$ decays) 
        to $10.7$ (including $W$ decays).
  \item The constraints on the fourth generation PMNS matrix elements obtained 
        from our study allow to draw several interesting conclusions: 
        \begin{itemize}
        \item In Ref.~\cite{Breskvar:2006fy} a unification of spins and charges has been 
              considered to explain family replication. A specific symmetry breaking mechanism 
              has been discussed in~\cite{Breskvar:2006fy} leading to the following prediction 
              for the fourth generation PMNS matrix:
              \begin{equation}
              \label{Eq:SpinChargeUnification}
              U_{PMNS} = 
              \left(
              \begin{array}{cccc}
               0.697 & 0.486 & 0.177 & 0.497 \\
               0.486 & 0.697 & 0.497 & 0.177 \\
               0.177 & 0.497 & 0.817 & 0.234 \\
               0.497 & 0.177 & 0.234 & 0.817
              \end{array}
              \right).
              \end{equation}
              By only considering the PMNS elements of the fourth row and column we conclude
              that our analysis clearly rules out the PMNS matrix predicted by this specific 
              symmetry breaking mechanism within the unification approach of spin and charges.
        \item In typical see-saw models~\cite{Weinberg,Fritzsch,FramptonHungSher} one expects 
              that the mixing angle between the fourth and the third generation $\theta_{34}$ 
              fullfils the relation $\sin^{2}{\theta_{34}} \approx m_{\tau}/m_{E}$ where 
              $m_{\tau}$ is the $\tau$-lepton mass and $m_{E}$ the mass of the fourth-generation 
              charged lepton. Using the LEP search limit, 
              $m_{E} > 100~{\rm GeV}$ at $90~\%$ CL~\cite{LEPLimitChargedLepton}, one finds 
              $\sin{\theta_{34}} < 0.13$. Since $\sin{\theta_{34}} \approx |U_{\tau 4}|$
              (in the Botella-Chau parametrisation~\cite{BotellaChau})
              our findings are in agreement with such a scenario if $m_{E}$ is at least of 
              order $240~{\rm GeV}$.
        \item While each individual system (leptonic $\tau$ decays, $K_{\ell 3}$ decays, and 
              leptonic $\pi$ decays) shows no signifcant deviation from lepton universality 
              between electrons and muons it is interesting to note that the deviation seen 
              in each case goes into the same direction and points to a non-zero value for 
              $|U_{e 4}|$. It should be stressed though that neither the individual nor the 
              combined constraints are sufficient to claim evidence for a significant deviation 
              of $|U_{e 4}|$ from zero. If confirmed by more stringent tests of lepton 
              universality this would point to New Physics. In case of a fourth generation it 
              would imply an interesting hierarchy: $|U_{e 4}| > |U_{\mu 4}|$ with a surprisingly 
              large value for $|U_{e 4}|$. To this end we note that the preliminary world average 
              $R_{K}=(2.498 \pm 0.014)~\%$~\cite{NA62Talk} for leptonic Kaon decays, if confirmed, 
              would reduce the observed deviation of $|U_{e 4}|$ from zero.
        \item In Ref.~\cite{HouMa} it was studied whether a heavy fourth generation neutrino
              could explain the discrepancy between the measured and predicted value of the 
              anomalous magnetic moment of the muon. It was concluded that this could be 
              only the case if the PMNS matrix element $|U_{\mu 4}|$ were of order $0.7$.
              The authors consider such a large value unrealistic as it has to fulfill
              the search limit $BF(\mu \rightarrow e \gamma) < 1.2 \times 10^{-11}$~\cite{LimitMuEGamma}
              which in turn requires the element $|U_{e 4}|$ to be very tiny at the same time.\\
              Our analysis which takes into account radiative and leptonic $\mu$ and 
              $\tau$ decays allows us now to draw a firm conclusion. Using the calculated
              contribution from a heavy fourth generation neutrino as given in Ref.~\cite{HouMa} 
              one is not able to explain the discrepancy between the measured and predicted value 
              for the anomalous magnetic moment of the muon since we find $|U_{\mu 4}|<0.029$ 
              ($2\sigma$ level) which is much smaller than the required size of O(0.7).
        \end{itemize}
\end{itemize}

\section{Summary}
Recent analyses of elements of the neutrino mass and quark mass mixing matrices as well as 
of electroweak precision fits within the framework of a fourth generation implicitely assume 
$3 \times 3$ unitarity for the first three generations in the neutrino sector or at least 
assume no mixing between the fourth and first two neutrino generations.
When relaxing this ad-hoc assumption we find that the value of the Fermi constant $G_F$ from 
the $\mu$-lifetime measurement can only be extracted with low precision and only when using 
also leptonic $\tau$ branching fractions. However, the leptonic $\mu$ and $\tau$ decays test 
lepton universality and thereby lead to strong correlations between the fourth generation 
PMNS matrix elements $|U_{e 4}|$, $|U_{\mu 4}|$, and $|U_{\tau 4}|$.
The precision on $G_F$ can only be improved significantly if one adds the search limits from 
radiative $\mu$ and $\tau$ decays which in turn allows to set interesting constraints on the
individual elements $|U_{e 4}|$, $|U_{\mu 4}|$, and $|U_{\tau 4}|$. Still, the precision on 
$G_F$ is at least two orders of magnitude worse compared to the 3SM value. $K_{\ell3}$ decays 
and leptonic decays of charged pions provide similar stringent tests of lepton universality 
and thereby improve further the constraints on $|U_{e 4}|$, $|U_{\mu 4}|$, $|U_{\tau 4}|$, 
and on $G_{F}$.\\ 
In our generalized analysis we find that the CKM sector is not very much affected with one
exception: the precision on $|V_{ud}|$ extracted from super-allowed $\beta$ decays is reduced.
This reduction in precision will impact the constraints on fourth generation CKM elements
imposed by $4 \times 4$ unitarity.
\\
Interestingly, the limits obtained on $|U_{e 4}|$, $|U_{\mu 4}|$, and $|U_{\tau 4}|$ are 
able to exclude already certain models that predict patterns for a fourth generation neutrino 
mixing matrix. In this context, it is interesting to note that we observe a p-value of 
$2.6~\%$ for $|U_{e4}|=0$. If this deviation could be confirmed with more precise tests 
of lepton universality it would indicate New Physics and possibly point to the existence 
of a fourth generation although such a large mixing between the fourth and first generation 
looks surprising.\\
We do not observe any violation of lepton universality between the $\tau$ and the $\mu$ 
sector in contrast to an evidence of such as found by \babar in a recent measurement of 
$BF(\tau \rightarrow h \nu_{\tau})$ ($h=\pi, K$) when compared to $BF(h \rightarrow \mu \nu_{\mu})$.\\
With $|U_{\mu 4}|<0.029$ found in our analysis we are able to exclude, using the study of
Ref.~\cite{HouMa}, the possibility that the discrepancy between the measured and predicted 
value for the anomalous magnetic moment of the muon can be explained by a contribution of 
a heavy fourth generation neutrino. \\
Future studies of fourth generation parameter constraints (PMNS matrix, CKM matrix and 
electroweak precision fit) need to relax the ad-hoc assumption of $3 \times 3$ unitarity 
for the first three generations. The authors are currently preparing studies within the 
CKMfitter package that combine the PMNS sector, the CKM sector, and the electroweak 
precision fit in a consistent way.

\section*{ACKNOWLEDGEMENTS}
This work has been performed within the CKMfitter group and we are grateful for the 
support provided by our colleagues. We thank M.~Vysotsky, M.~G\"obel (Gfitter group),
M.~Antonelli (Flavianet) for communications, and G.~Kribs, T.~Plehn and M.~Spannowsky 
for clarifications. We also thank R.~E.~Shrock, E.~Fernandez-Martinez and B.~Gavela 
for pointing us to earlier, related work. We appreciate the enlightening discussions 
with U.~Nierste and A.~Lenz on the document.

\end{document}